\documentclass[twocolumn,superscriptaddress,showpacs]{revtex4}
\usepackage{graphicx}
\usepackage{setspace}
\usepackage{titlesec}
\usepackage{epsfig}
\usepackage{amssymb}

\begin{document}

\title{Design of magnetic traps for neutral atoms \\ with vortices in type-II superconducting micro-structures} 


\author{B.~Zhang}
\affiliation{Division of Physics and Applied Physics, Nanyang Technological University,  21 Nanyang Link, Singapore 637371}
\author{R.~Fermani}
\affiliation{Division of Physics and Applied Physics, Nanyang Technological University,  21 Nanyang Link, Singapore 637371}
\affiliation{Centre for Quantum Technologies, National University of Singapore, 3 Science Drive 2, Singapore 117543}
\author{T.~M{\"u}ller}
\affiliation{Division of Physics and Applied Physics, Nanyang Technological University,  21 Nanyang Link, Singapore 637371}
\affiliation{Centre for Quantum Technologies, National University of Singapore, 3 Science Drive 2, Singapore 117543}
\author{M.~J.~Lim}
\affiliation{Division of Physics and Applied Physics, Nanyang Technological University,  21 Nanyang Link, Singapore 637371}
\affiliation{Department of Physics and Astronomy, Rowan University, 201 Mullica Hill Road, NJ, USA}
\author{R.~Dumke}
\email{rdumke@ntu.edu.sg}
\affiliation{Division of Physics and Applied Physics, Nanyang Technological University,  21 Nanyang Link, Singapore 637371}

\date{\today}

\begin{abstract}
We design magnetic traps for atoms based on the average magnetic field of vortices induced in a type-II superconducting thin film. This magnetic field is the critical ingredient of the demonstrated vortex-based atom traps, which operate without transport current. We use Bean's critical-state method to model the vortex field through mesoscopic supercurrents induced in the thin strip. The resulting inhomogeneous magnetic fields are studied in detail and compared to those generated by multiple normally-conducting wires with transport currents. Various vortex patterns can be obtained by programming different loading-field and transport current sequences. These variable magnetic fields are employed to make versatile trapping potentials. 
\end{abstract}


\maketitle 

\section{Introduction}

The development of microchips for trapping and manipulating ultracold atoms has progressed rapidly in the last decade~\cite{Hinds,Folman02,Fortagh07}. With lithography and other processes, complex surface structures can be fabricated to provide various tight trapping potentials for cold atoms in the proximity of the microchip surface. However, the fluctuating electromagnetic fields emanating from metallic surfaces may lead to loss and decoherence of trapped atoms, thus limiting the performance of the atom chip~\cite{Henkel99c,Jones,Harber} in devices such as atom interferometers and atomic clocks. Superconducting atom chips have been shown to reduce both the near field noise and technical noise~\cite{Skagerstam06a,Hohenester07,Mukai,Nirrengarten,Roux,Kasch,Fermani09,Scheel05a,Scheel07}.  The spin-flip lifetime and coherence of the trapped atoms improves accordingly. Additionally, superconducting atom chips may provide a platform to realize a coherent interface between atomic or molecular quantum states and quantum solid-state devices~\cite{Tian,Andre,Tordrup,Verdu}.  

A special property of type-II superconductors in the mixed state is the presence of vortices. 
The effect of vortices on the trapping potentials generated by current-carrying superconducting structures in the mixed state has been studied~\cite{Dikovsky,Emmert09}. Recently, stable magnetic traps without applied transport current that use the fields produced by vortices trapped in a superconducting strip~\cite{Mueller09} or disk~\cite{Shimizu} have been experimentally demonstrated. This suggests that vortices could provide a bridge between coherent atomic states and solid-state quantum devices, enabling novel applications and interesting new fundamental studies. The properties of these vortex-based traps depend directly on the vortex distribution.  This also implies that sufficiently cold atoms could be used as a sensitive probe of the distribution and dynamics of vortices in superconductors~\cite{Shimizu,Mueller09}.

In this paper, we describe in detail how to design magnetic traps for low-field seeking atoms using vortex penetration in a type-II superconducting thin film. We show how to create various trap types such as a single harmonic type trap, a double trap, or traps without the typically applied bias field, and we include experimental characterizations related to our theoretical findings. These traps form at distances large compared to the characteristic vortex-vortex spacing, so it is useful to model the vortex field with the Bean critical-state approximation \cite{Bean64,Brandt}.  This approach replaces the individual vortices in a mixed state (Shubnikov state) superconductor with an equivalent current density in a finite-width thin strip, which accurately describes our setup geometry.  The mesoscopic equivalent supercurrent can be induced either by an external magnetic field pulse perpendicular to the surface or by a transport current pulse through the superconductor; we study both cases.  Straightforward application of the Biot-Savart law then yields the potential relevant for an atom with a magnetic moment that is in the vicinity of the microchip.  This computation has enabled the discovery and optimization of several vortex-based trap geometries that we have subsequently observed in experiments~\cite{Mueller10}: a micro-trap resembling the Z-type wire-trap geometry, a double trap, and a self-sufficient trap requiring no external bias field. While analogous potentials have been demonstrated using normally-conducting wires, we expect that in our superconducting atom chip (during the atom-trapping phase), the near-field noise is significantly reduced and technical noise decreases as the transport current is absent. 

The paper is organized as follows. Sec.~\ref{sec:flux} contains a brief summary of magnetic flux penetration in type-II superconductors. In Sec.~\ref{sec:B} we study the mesoscopic equivalent supercurrents induced by a pulsed external field perpendicular to the strip surface. The induced supercurrent distribution is computed along with the resulting external magnetic field, which is the basis for the magnetic traps. In Sec.~\ref{sec:I} we provide similar analysis for supercurrent distributions that remain after applying a pulsed transport current. In Sec.~\ref{sec:discuss} we explore the impact of the transport current history on the magnetic atom trap and on the trapped vortices. Concluding remarks are given in Sec.~\ref{sec:con}.

\section{Magnetic flux penetration}
\label{sec:flux}
We briefly review magnetic flux penetration in the form of quantized vortices, in order to introduce the physics background relevant to the simulations presented in the following sections. 
Vortex penetration exists for only type-II (as opposed to type-I) superconductors. The order parameter of a superconductor  is given by $\kappa=\lambda/\xi$, where $\lambda$ is the London penetration depth and $\xi$ is the coherence length. Type-I superconductors have $\kappa < 1/\sqrt{2}$ and their transition from the superconducting phase to the normal conducting state occurs for external magnetic fields $B_{ext}  $ higher than the thermodynamic critical field $ B_c (T)= 
\Phi_0 / \sqrt{8} \pi \xi(T) \lambda (T)$, where $\Phi_0= h/2e= 2.07 \, 10^{-15}$ Tm$^2$ is the quantum of magnetic flux. For an applied field $B_{ext}< B_c (T)$, the superconductor is in the Meissner state and flux penetrates only into a thin surface layer of depth $\lambda$. In this layer, the shielding currents can reach the critical current density $j_c (T)= B_c(T)/\mu_0 \lambda(T)$.
In contrast, Type-II superconductors have $\kappa > 1/\sqrt{2}$ and are characterised by two critical magnetic fields: $B_{c1}(T) \simeq \Phi_0/ 4 \pi \lambda(T)^2 \leq B_c(T)$ below which the superconductor is in the Meissner state, and $B_{c2} (T)= \Phi_0/ 2 \pi \xi(T)^2 \geq B_c(T)$ above which the superconductor is in the normal conducting state. In the field range $B_{c1}(T)< B_{ext} < B_{c2} (T)$, the superconductor is said to be in the mixed (or Shubnikov) state and magnetic flux penetrates in the form of vortices. Each single vortex carries a quantum of magnetic flux and consists of a normal core of radius $\approx \xi$ around which shielding currents flow within a radius $\sim \lambda$~\cite{Kleiner}. 

In an ideal type-II superconductor without defects or fluctuations, vortices enter the sample from the edges and arrange themselves in a triangular lattice (Abrikosov lattice).
The distance between vortices decreases when $B_{ext}$ and $T$ increase. Shielding currents with density $j_c$ flow in the penetrated region and are no longer restricted to a thin surface layer of thickness $\lambda(T)$, but can also flow in the interior of the sample. 
For temperatures close to $T_c$, the magnetic flux penetrates uniformly into the sample and can be described by means of a mesoscopic equivalent supercurrent as in the critical-state models~\cite{Brandt,Bean64}. At temperatures below an activation temperature $T_a<T_c$, the flux penetration happens via dendritic avalanches that propagate into the sample from the edge~\cite{Mints}. However, the study presented in this paper assumes that the superconductor temperature is close to $T_c$, so magnetic flux penetration is accurately described by the models given in~\cite{Brandt,Bean64}.

Actual (non-ideal) type-II superconductors have defects that act as pinning centers where vortices are more likely to form. The pinning energy competes with the mutual repulsion of the vortices, so an energetically unfavorable arrangement of vortices will change because of thermal fluctuations. Thermal fluctuations may lead to the depinning of single vortices by providing the flux line enough energy to leave the pinning center. During the application of an external magnetic field, an equilibrium between the pinning energy and the vortex-vortex repulsion is reached, and the resulting vortex distribution may resemble a glass phase without long-range spatial order. After the magnetic field is removed, a new equilibrium configuration must be reached. The vortices remaining in the sample produce the remanent magnetization and the superconductor is said to be in the remanent state. 
The vortex configuration in the remanent state can be subsequently modified by applying a transport current. The transport current passes through the vortices and a Lorentz force arises between the electric current and magnetic flux lines. As a consequence, some vortices may leave the superconductor or accumulate asymetrically with respect to their initial equilibrium configuration~\cite{Kleiner}. 

In previous experimental investigations~\cite{Mueller10}, we measured the location of the vortex-based magnetic atom trap to be stable over $\sim 2$  hours.  This confirms that any thermal relaxation of the vortices does not affect the overall macroscopic magnetization of the superconductor on the experiment timescale.  We assume that the vortex configuration is stable as long as no transport current is applied.

\section{Magnetization by external magnetic field}
\label{sec:B}
\begin{figure}[h]
 \includegraphics[width=7cm]{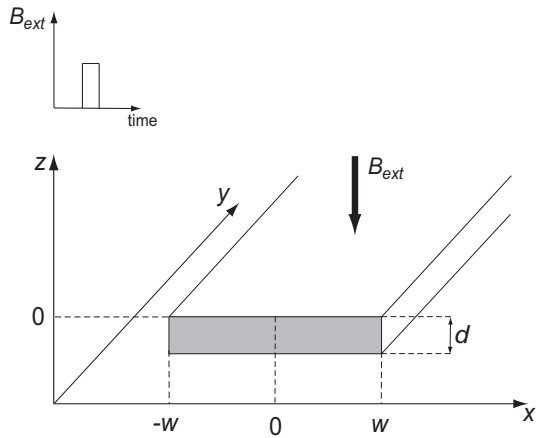} 
 \caption{\label{fig:skatch} Schematic of the superconducting strip where $d$ is the strip thickness and $2 w $ the strip width. The strip is assumed to be infinitely long in the $y$-direction.  A magnetic field pulse $B_{ext} $ is applied perpendicularly to the strip surface. When the amplitude is such that $B_{c1} < B_{ext} < B_{c2}$, magnetic flux penetrates the strip from the edges, i.e.~from $x=\pm w$.}
 \end{figure}

In this section, we consider a magnetic field $B_{ext}$ applied perpendicularly to a superconducting strip as shown in Fig.~\ref{fig:skatch}. We assume that the strip thickness $d$ is much smaller than the width $2w$ and that the strip is infinitely long in the $y$-direction, such that we can reduce our study to the $xz$-plane. For magnetic field amplitudes such that $ B_{c1} < B_{ext} < B_{c2} $, the magnetic flux enters the strip from the edges in the form of vortices.
The average magnetic field of the vortices can be accounted for by computing the magnetic field generated by the equivalent mesoscopic supercurrents in the strip. In the following sections we study the vortex field only through the supercurrent distribution. 
Due to the memory effect of the superconductor, supercurrents remain in the strip even after the external field is completely removed. We present several magnetic trap configurations relative to various loading field sequences.

\subsection{Induced supercurrent distribution and field distribution}
\label{sec:Bcurrent}

We simulate the current distribution for an external magnetic field $ B_{c1} < B_{ext} < B_{c2} $ applied perpendicularly to the type-II superconducting thin strip in the virgin state. In the regions where flux penetrates, the current density reaches the critical value $j_c$, according to Bean's critical-state model~\cite{Bean64,Brandt}. As  $d \ll 2w$, the current density in the strip is assumed to be constant over the strip thickness $d$. For simplicity, to model the supercurrent distribution we use a sheet current density defined as $J(x)=j(x) d$, where $j(x)$ is the local current density. The sheet current density $J(x)$ in the strip is given by~\cite{Brandt} 
\begin{equation}
\label{eq:Brandt-B}
J(x,B_{ext}, J_c) = \!\left \{ \begin{array}{ll}
\frac{2J_{c}}{\pi}\arctan \left ( \frac {x}{w} \sqrt{\frac{w^2-b^2}{b^2-x^2}} \right )  , \,&  |x| \le b \\
J_{c} \frac{x}{|x|} \; ,    & b \le |x| \le w
\label{eq:Jvirgin}
\end{array} \right.
\end{equation} 
where $b=w/\cosh(B_{ext}/B_{c})$ denotes the half width of the central flux-free region, $B_{c}=\mu_0 J_{c} / \pi$ is the thermodynamic critical field, $J_c=j_c d$ is the critical sheet current density and $2w$ is the strip width. The width of the outer region penetrated by the vortices is given by $w-b$. The maximum penetration field $B_{p}$ of the strip, which can be described by Brandt's model~\cite{Brandt}, is obtained when $b$ goes to zero, yielding
\begin{equation}
 B_{p}= \left (  \frac{ \mu_0 j_c d}{\pi} \right ) \ln \left (\frac{2w}{d}\right ) \; . \label{eq:upper}
 \end{equation}

\begin{figure}
\begin{center}
\includegraphics[width=6cm]{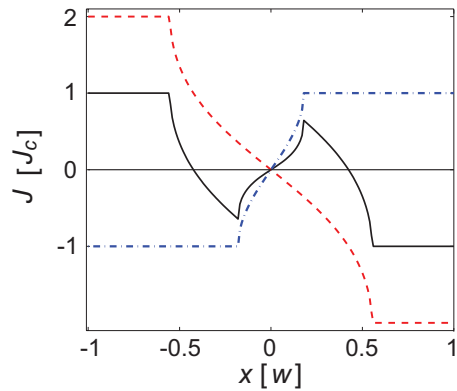} 
\end{center}
\vspace{-0.4cm}
\caption{(color online) Distribution of the sheet current density across the strip for $B_{ext}= 0 \to 2.4 B_{c}  \to 0$. Blue dot-dashed line: sheet current density for $B_{ext}= 0 \to2.4 B_{c}$; red dashed line: sheet current density for $B_{ext}= 2.4 B_{c} \to 0$; black solid line: final sheet current density. The $x$-position is normalized by $w$, and the sheet current density is normalized by $J_c$.  \label{fig:current} }
\end{figure}

When $B_{ext}$ decreases to its final value, $B_{f}$, and $-B_{ext}<B_{f}<B_{ext}$, the sheet current density can be described by~\cite{Brandt}
 \begin{equation}
 J(x, B_{f}, J_c)=J(x,B_{ext}, J_c)-J(x,B_{ext}-B_{f}, 2J_c) \; . \label{eq:Jdown}
 \end{equation}
We plot in Fig.~\ref{fig:current} the sheet current density distribution resulting from the application of a field pulse $B_{ext} = 2.4 B_{c}$.  An increase of $B_{ext}$ from zero to $2.4 B_{c}$ induces a sheet current density $J(x,2.4 B_{c}, J_c)$ computed by Eq.~(\ref{eq:Jvirgin}) and represented by the blue dot-dashed line in Fig.~\ref{fig:current}. When $B_{ext}$ is decreased to zero, some anti-vortices penetrate the strip from the edges. This is described by a sheet current density $J(x,0, 2J_c)$  with opposing sign with respect to $J(x, 2.4 B_{c}, J_c)$, and it is represented by the red dashed line Fig.~\ref{fig:current}. The final current density is the sum of these two curves as in Eq.~(\ref{eq:Jdown}), and it is represented by the black solid line in Fig.~\ref{fig:current}. For symmetry reason, the supercurrents flow in opposite directions in the two halves of the strip and the net current in the entire strip is zero. Most supercurrents flow in the regions given by $0.4w<|x|<w$; the supercurrents in the central part are considerably smaller. We integrate the final sheet current density $J(x)$ over one half of the strip and find that the current magnitude is $\pm 0.37j_c d w$. In simulations throughout this paper we will always consider a sample strip with $2w=400 \; \mu$m width, $d=1 \; \mu$m thickness and $j_c = 2.08 \times 10^6$~A/cm$^2$ resulting in a maximum penetration field $B_p = 499.27$~G and a thermodynamic critical field $B_c=\mu _0 J_c/\pi = 83.33$~G.  For example, the application of a field pulse of $B_{ext}=200$~G induces a supercurrent of magnitude $\pm1.55$~A in each half of the strip. The superconducting strip thus resembles two parallel wires with counter-propagating currents. 

 \subsection{Design of magnetic trap}
\label{sec:Btrap}

\begin{figure}[h]
\begin{center}
\includegraphics[width=6.cm]{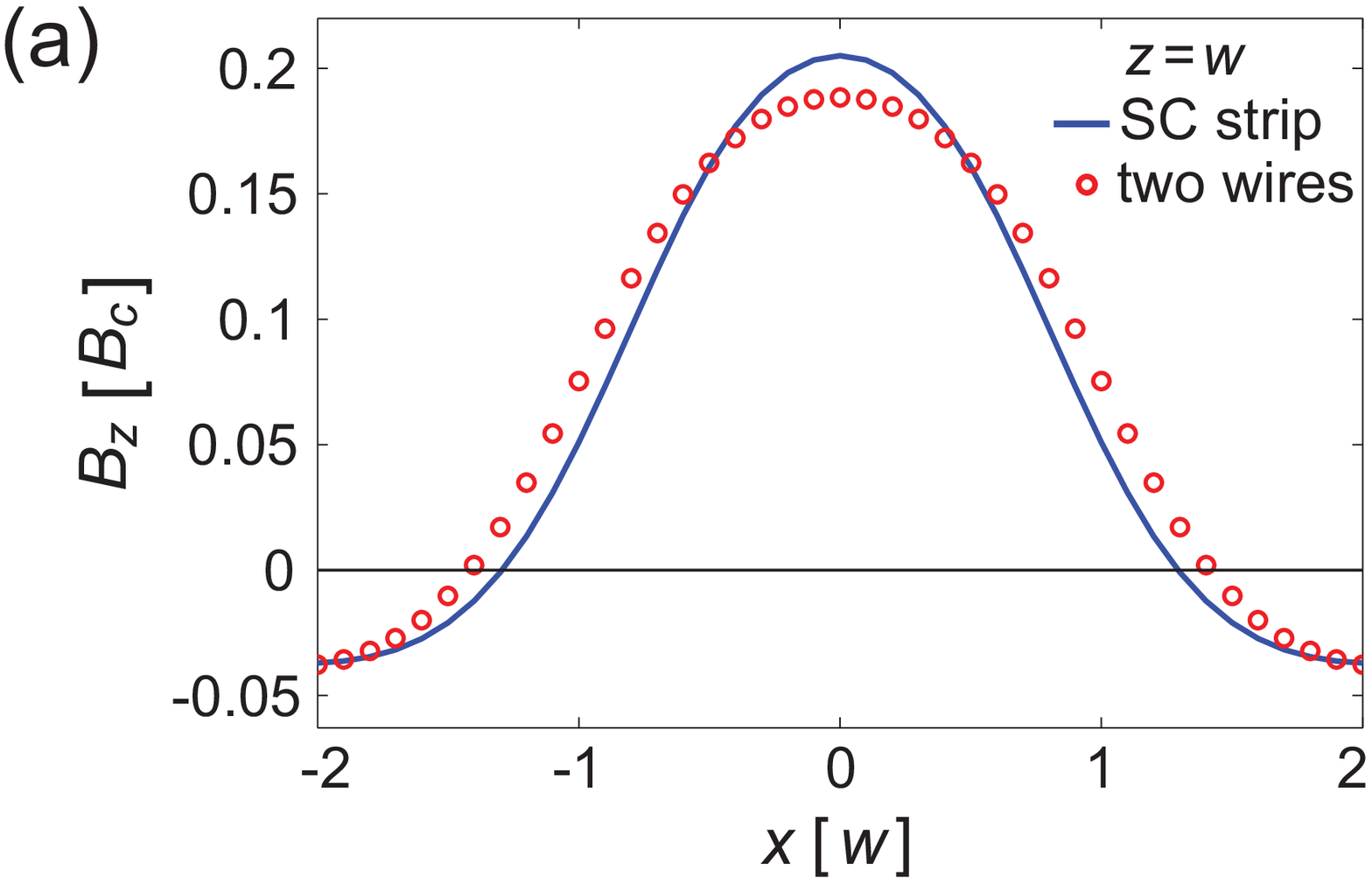} 
\\ \vspace{0.2cm}
\includegraphics[width=6cm]{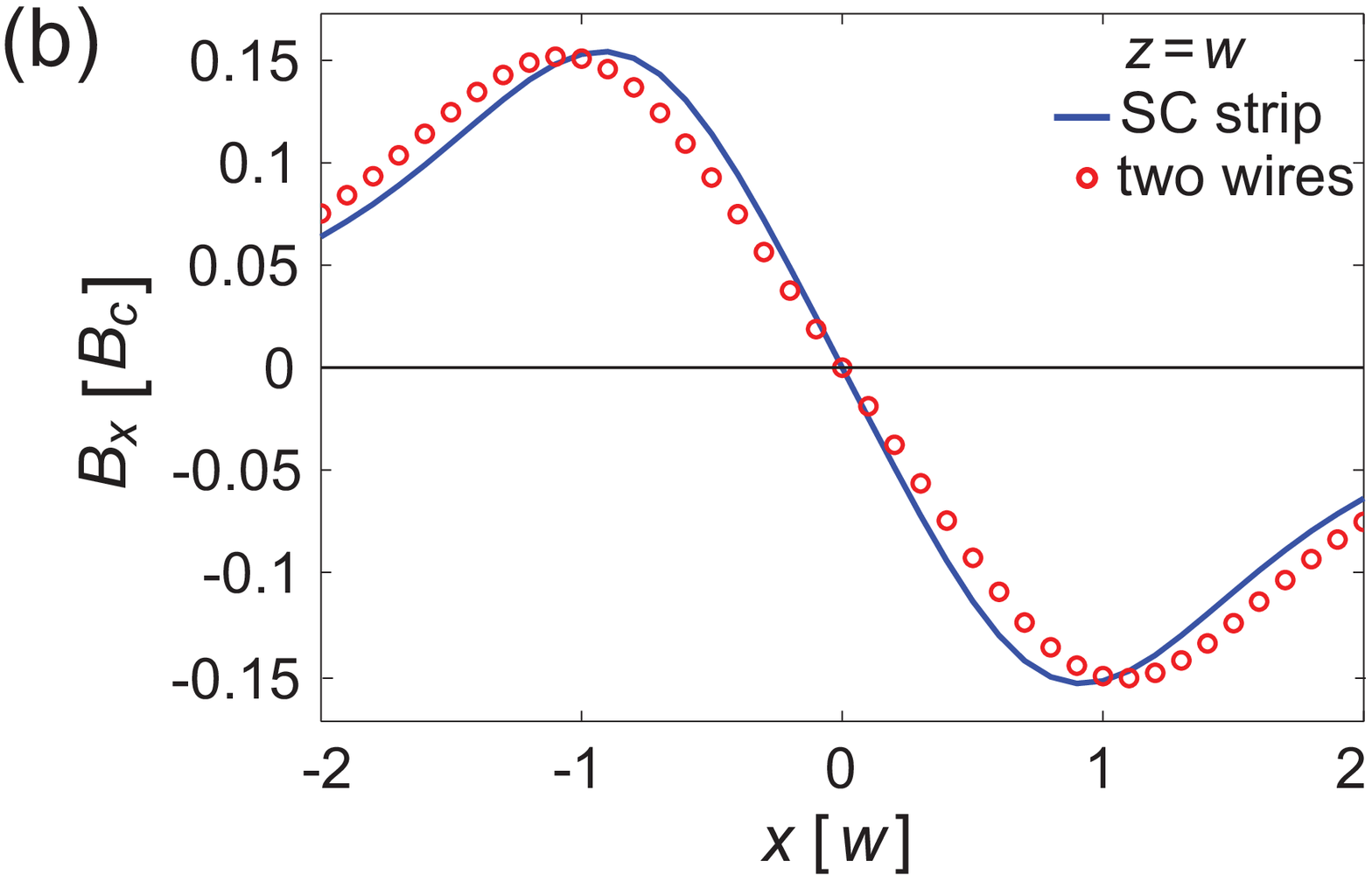}
\\ \vspace{0.2cm}
\includegraphics[width=6cm]{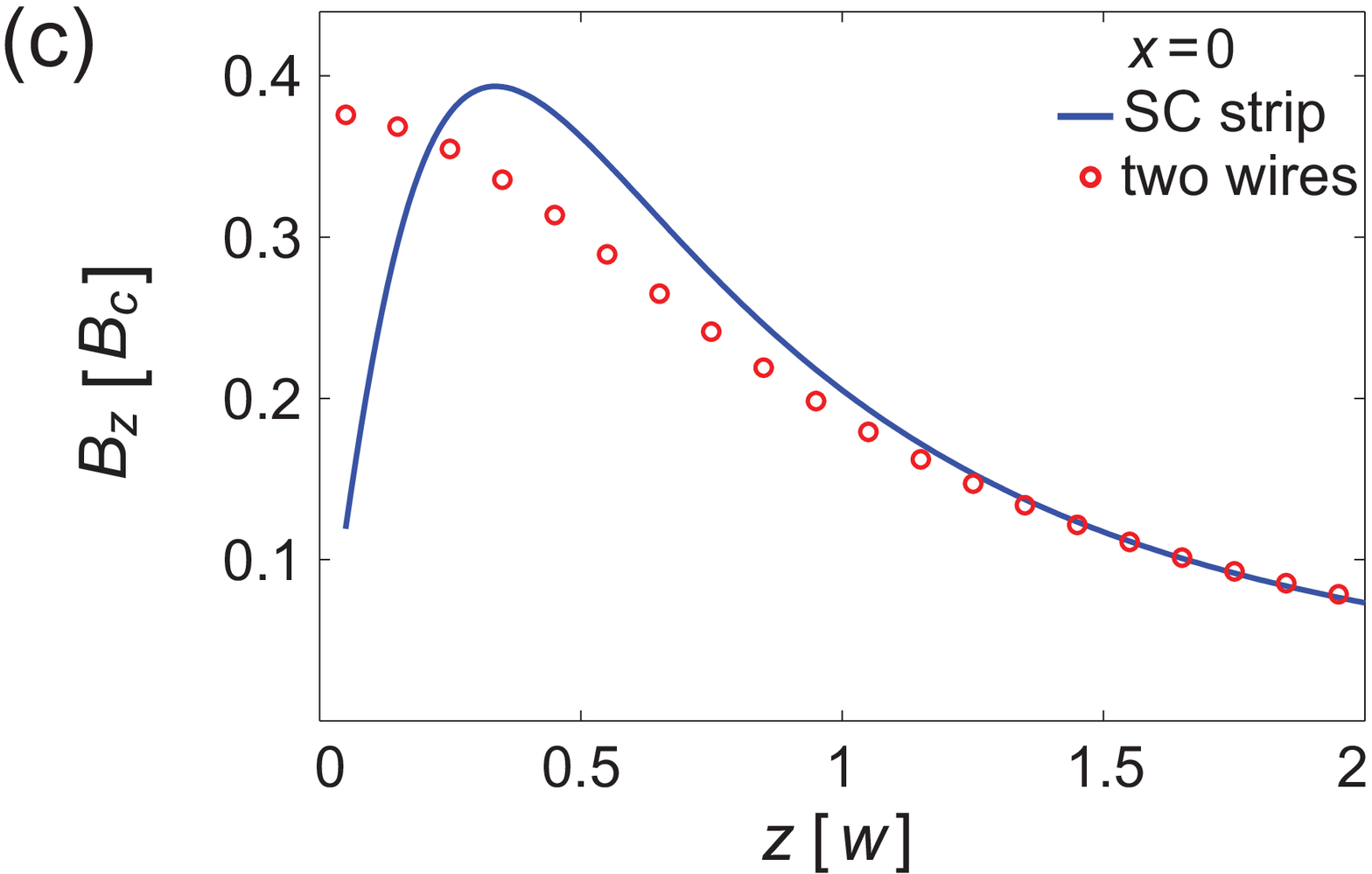}
\end{center}
\vspace{-0.5cm}
\caption{(color online) Components of the magnetic field generated by equivalent supercurrents relative to  $B_{ext}= 0 \to 2.4 B_{c}  \to 0$ (blue lines) and by two counter-propagating currents $ \pm 0.37j_c d w$  (red circles). (a) $B_z(x,w)$ vs $x$ at fixed height $z=w$;  (b) $B_x(x,w)$ vs $x$ at fixed height $z=w$; (c) $B_z(0,z)$ vs $z$ above the strip center $x=0$. The $x$- and $z$- positions are normalized by $w$ and the magnetic field is normalized by $B_c$. \label{fig:Bvortex}}
\end{figure}

We compute the magnetic fields generated by the supercurrents (equivalent to the average field of all vortices) by applying the Biot-Savart theorem. We plot the $z$-component of the vortex field, $B_z(x,w)$, along the $x$-axis in Fig.~\ref{fig:Bvortex}(a) at a distance $z=w$. Note that this component is zero at two points near the strip edges where quadrupole-type confinement can be generated by cancelling the $x$-component of the vortex field with a bias field, we discuss this trap configuration in Sec.~\ref{sec:Bpara}.  We plot in Fig.~\ref{fig:Bvortex}(b) the $x$-component of the vortex field, $B_x(x,z)$, along the $x$-axis at a distance $z=w$, for symmetry reasons this component is always zero above the strip center, $B_x(0,z)=0$.  The $z$-component of the vortex field $B_z(0,z)$ is nonzero above the strip center, and we plot $B_z(0,z)$ as a function of $z$ in Fig.~\ref{fig:Bvortex}(c). Therefore,  by cancelling the $z$-component of the vortex field above the strip center, a field minimum can be obtained to confine low-field seeking atoms. This is discussed in detail in  Sec.~\ref{sec:Bperp}. 
 For comparison, we also plot the magnetic fields created by two counter-propagating currents in parallel, normally-conducting wires in Fig.~\ref{fig:Bvortex} (red circles). The two wires are at $x=\pm w$ and carry a current of $ \pm 0.37j_c d w$.  The superconducting strip and the two current-carrying wires give rise to similar spatial magnetic field distributions. Only $B_z(0,z)$ is qualitatively different at short distances $z<w$ as shown Fig.~\ref{fig:Bvortex}(c), because a superconductor expels all magnetic fields perpendicular to its surface. In the following, we discuss in detail various trapping potentials with these magnetic fields.  

\subsubsection{Magnetic trap with parallel bias field}
\label{sec:Bpara}

\begin{figure}[t]
\begin{center}
\includegraphics[width=5.8cm]{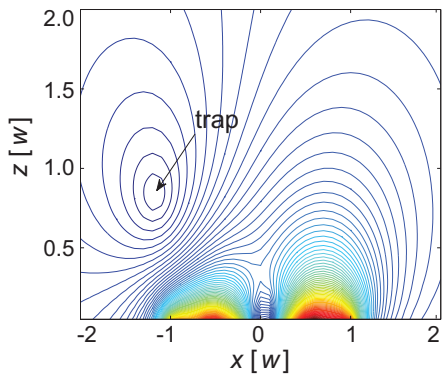}
\end{center}
\vspace{-0.6cm}
\caption{(color online) A quadrupole type confinement above the edge of the strip formed by the vortex field together with a parallel homogeneous field $B_{bias}^x= -0.173B_c$. The vortex-loading field pulse has an amplitude of $2.4 B_{c}$. $x$- and $z$-positions are normalized by $w$. \label{fig:Btrapp} }
\end{figure}

 A quadrupole trap can be formed above one of the two strip edges by applying a homogenous bias field $B_{bias}^x$ along the $x$-direction such that it cancels the $x$-component of the vortex field and $B_{bias}^{x}+B_x(x,z)=0$. For simplicity, we assume that the strip is infinitely thin, so $B_{bias}^x$ induces no current. The trap can be generated at various distances $z$ by changing the bias field $B_{bias}^x$ accordingly. This kind of trap has been recently observed experimentally in \cite{Mueller09} and we show a typical magnetic field configuration in Fig.~\ref{fig:Btrapp}. Such a configuration is formed with a field-loading pulse of amplitude $2.4B_c$ and a bias field $B_{bias}^x= -0.173B_c$, and the trap is formed at a distance $z=0.8w$. Simulations with our superconducting sample strip yield a field pulse of amplitude $2.4B_c=200$~G and a homogenous field $ B_{bias}^x = 14.4$~G generating quadrupole-type confinement at a height of $0.8w=160 \; \mu$m.\\

As shown in Fig.~\ref{fig:current}, there are counter-propagating supercurrents in each half of the strip. However, the currents in the central part of the strip are considerably smaller than in the outer region and the entire superconducting strip resembles two normally-conducting wires with counter-propagating currents.
The current near the edges gives the dominant contribution to the magnetic field distribution and the magnetic field at distances $z<w$ above the strip edges resembles the magnetic field generated from  a single current-carrying wire.   
 
 Our simulations consider only the two dimensional confinement, leading to a guiding potential along the wire in $y$-direction as in Fig.~\ref{fig:Btrapp}. However, to realize actual trapping of atoms, confinement along the
  $y$-direction must be provided as well. In our experimental realization of the described trap, this axial confinement is due to patterning the strip in a Z-shape~\cite{Mueller09}. In a simplified schematic, this geometry is similar to two normally-conducting Z-wires carrying opposing currents. Following this analogy, we employ only one of these wires and the magnetic field along $y$ originating from the corners of the Z-shape provides axial confinement. The axial field is due to the curvature of the magnetic field component $B_y$, and has a non-vanishing value at the center of the quadrupole-type radial potential. In total, this leads to finite magnetic field and a quasi-harmonic potential near the trap minimum. In our actual apparatus, similar magnetic fields arise from the corners of the vortex-loaded, Z-shaped superconducting strip.  This produces confinement along $y$ for all of the traps presented in this work. However, all subsequent discussion of potentials is limited to two dimensional confinement in the $xz$-plane.
 
\begin{figure}[b]
\begin{center}
\includegraphics[width=6.8cm]{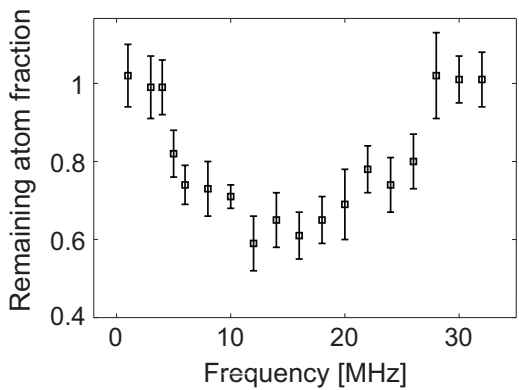}
\end{center}
\vspace{-0.6cm}
\caption{Atom loss spectrum of the Z-type micro-trap.  The remaining fractional population is determined by comparing resonant laser absorption images with and without a previous $50$~ms rf pulse; error bars reflect statistical uncertainty.  The field value at the trap minimum is $6.6$~G, which offsets the low-frequency threshhold to approximately $4$~MHz.  The inferred trap depth is about $1$~mK. \label{fig:rf} }
 \end{figure}
In our experimental realization, we also apply an offset field $ B_{bias}^y$ along the $y$-direction to increase the absolute magnetic field at the trap minimum. This is done to reduce Majorana spin-flip losses.  We characterize this trap by performing rf atom-loss spectroscopy to determine the depth of the magnetic potential. We load atoms into the trap, apply a $50$~ms rf pulse, then use absorption imaging to determine the remaining population.  This number is compared to a background image (no applied rf) to determine the remaining fractional population.  In Fig.~\ref{fig:rf} the remaining fraction as a function of applied frequency is shown.
The minimum trap field value has been offset with a $6.6$~G uniform bias field pointing along the $y$-axis of the trap, which is why the loss threshold occurs at approximately $4$~MHz rather than near zero frequency.  The $23$~MHz frequency interval over which stretched-state rubidium atoms are removed from the trap implies a potential well-depth on the order of $1$~mK. 

\begin{figure}[h]
\begin{center} 
\includegraphics[width=7cm]{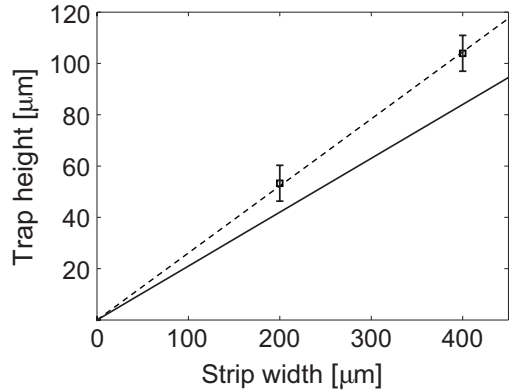}
\end{center}   
\vspace{-0.6cm}
\caption{Experimentally determined microtrap-to-surface distance scaling with strip width $w$. The trap-to-surface distance is plotted for a fixed horizontal bias field of $26$ G and superconductor strip widths of $200 \; \mu$m and $400 \; \mu$m.  Error bars reflect systematic uncertainty, and the dashed curve is a linear fit constrained to intersect the origin.  Adequate agreement is found with the prediction (solid curve) from a simulation based on the Brandt model.  The linear relationship arises from the direct proportionality of the remanent-state current to the width of the superconducting strip. \label{fig:zvsw}}
 \end{figure}

The superconducting chip employed in our experiments features two micro-structures with different width. Our simulations predict linear scaling of the trap-to-surface distance with the strip width, a consequence of the direct proportionality of the remanent-state sheet current to the strip width. We experimentally observe atoms in the described micro-trap type above both these strips. For this measurement both strips have been prepared using the same applied magnetic fields. We show the measured trap-to-surface distance for the different widths in Fig.~\ref{fig:zvsw}. 
The measurement confirms the predicted scaling, as shown by the dashed curve, which is a linear fit constrained to go through the origin. The solid curve reflects the proportionality constant from our simulations.  Although the measurements do not overlap with the theory, this minor difference is consistent with other types of measurements that previously revealed discrepancies with the Brandt model~\cite{Bean64,Brandt}. The relative agreement indicates the validity of the simplifications made in the model.

\subsubsection{Magnetic trap with perpendicular bias field}
\label{sec:Bperp}

\begin{figure}[h]
\begin{center}
\includegraphics[width=5.8cm]{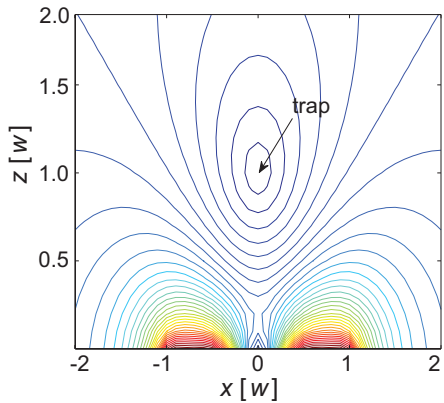}
\end{center} 
\vspace{-0.5cm}
\caption{(color online) A quadrupole trap above the strip center formed by the vortex field together with a perpendicular homogeneous field $B_{bias}^z= -0.2B_c$. The vortex-loading field pulse has an amplitude of $2.4 B_{c}$. The $x$- and $z$-positions are normalized by the strip width $w$. \label{fig:Btrapv} }
\end{figure}

A quadrupole-type confinement can be realized with a perpendicular bias field as well. We show in Fig.~\ref{fig:Bvortex}(b) that $B_x(0,z)$ is always zero above the strip center, while $B_z(0,z)\neq 0$ as in Fig.~\ref{fig:Bvortex}(c). The nonzero value of $B_z(0,z)$ can be cancelled with a homogeneous bias field $B_{bias}^{z}$ to generate a minimum  at $(0,z)$ in the magnetic potential. A conceptually similar trap has been experimentally demonstrated in~\cite{Shimizu}. The perpendicular bias field $B_{bias}^{z}$ also induces supercurrents in the strip, which we account for in our simulations. 
When the sum of the induced fields and the bias field is zero, a magnetic trap forms at $(0,z)$. 
A small perpendicular field, for example $B_{bias}^{z} \sim -0.2 B_c$, generates a quadrupole trap near $z=w$ above the center of the strip, as shown in Fig.~\ref{fig:Btrapv}. Note that there is a second trap directly on the surface at the strip center as $B_z(0,0)$ and $B_x(0,z)$ are always zero. The simulation done for our sample using a bias field $ B_{bias}^z = 16.6$~G, shows a quadrupole trap at a height of $200 \; \mu$m. 

\subsubsection{Magnetic trap without bias field}
\label{sec:Bself}
\begin{figure}[h]
\begin{center}
\includegraphics[width=5.6cm]{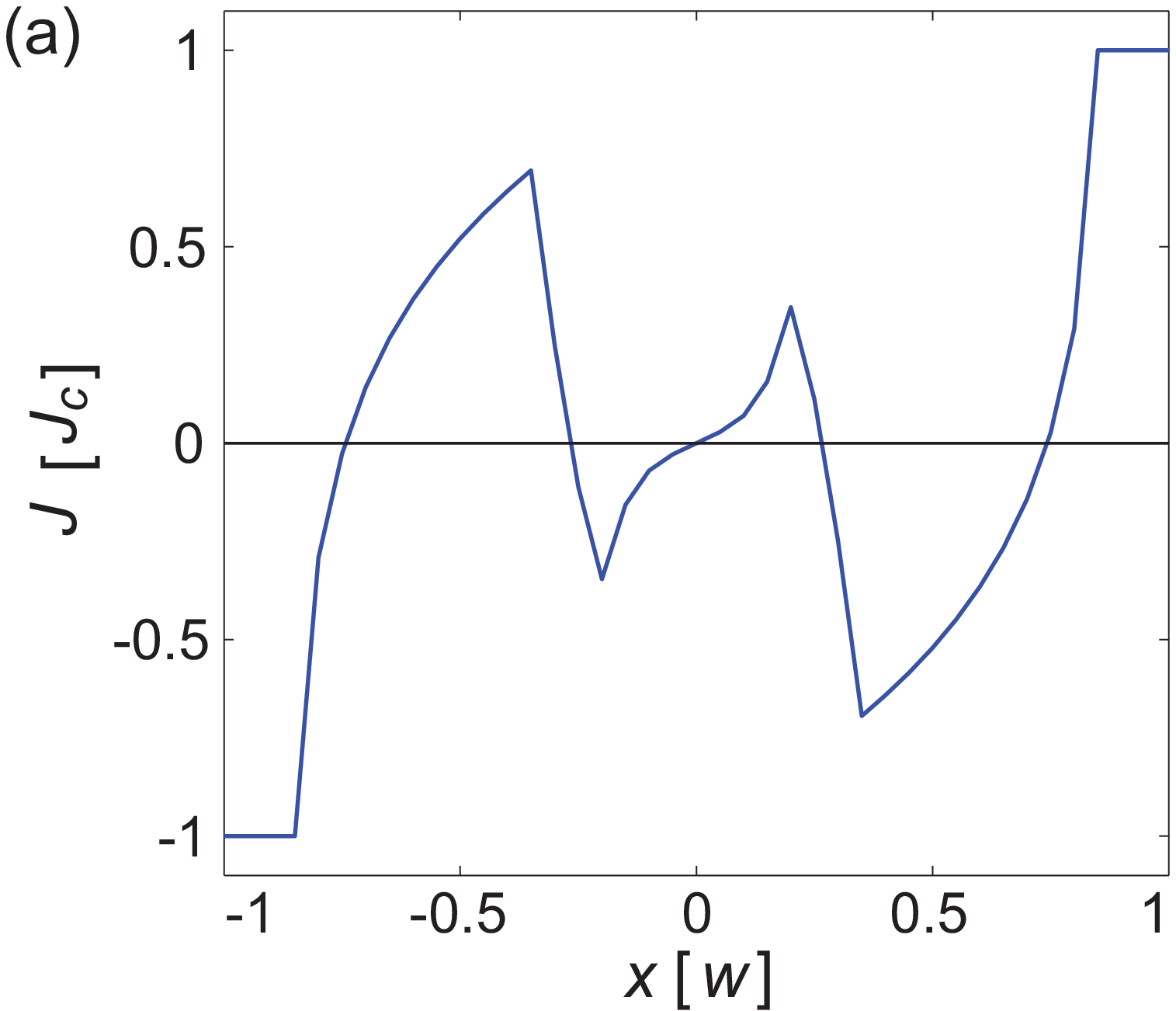}\\
\vspace{0.2cm}
\includegraphics[width=5.6cm]{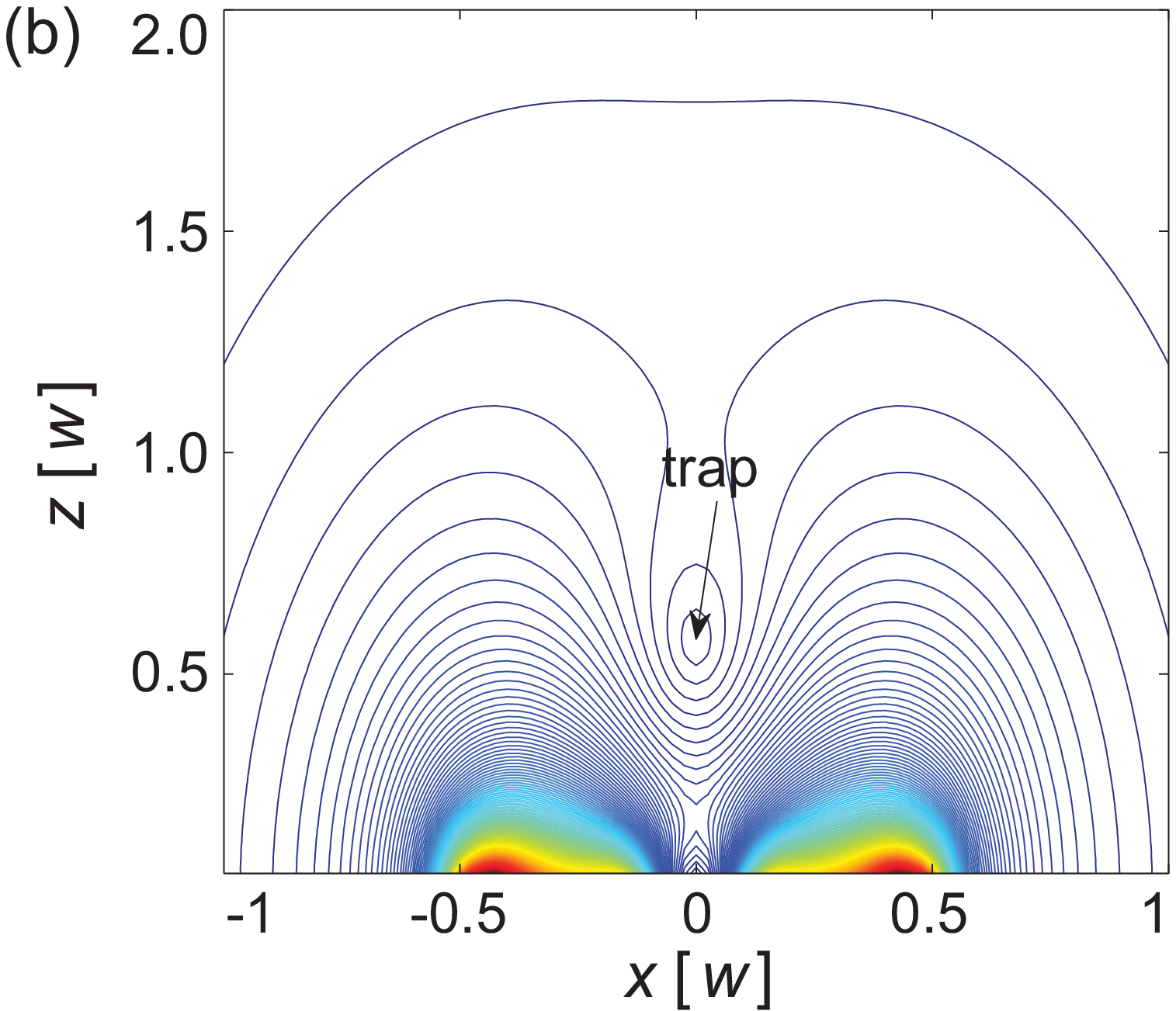}
\end{center}
\vspace{-6mm}
\caption{(color online) Loading field perpendicular to the strip with sequence $B_{ext}= 0\to 2.4B_c \to -1.2B_c \to 0$. (a)  current density distribution in the strip. (b) equipotential lines for an atom with nonzero magnetic moment.  A trap can be seen at a height of $z=0.6w$. \label{fig:Bvortextrap}  }
 \end{figure}

  The superconductor records the history of sign changes in applied magnetic fields. Sequences of field reversals can be used to induce complex current distributions allowing the creation of versatile magnetic field geometries. We present a specially designed supercurrent distribution generating a self-sufficient trapping field able to store atoms without any other bias fields. We consider a magnetic field perpendicular to the strip in the virgin state, increase the field from zero to $2.4B_c$  and then decrease to $-1.2B_c$, before finally removing the external field completely. The resulting current density distribution in the strip is plotted in Fig.~\ref{fig:Bvortextrap}(a). 

\begin{figure}[h] 
\begin{center} 
\includegraphics[width=5.6cm]{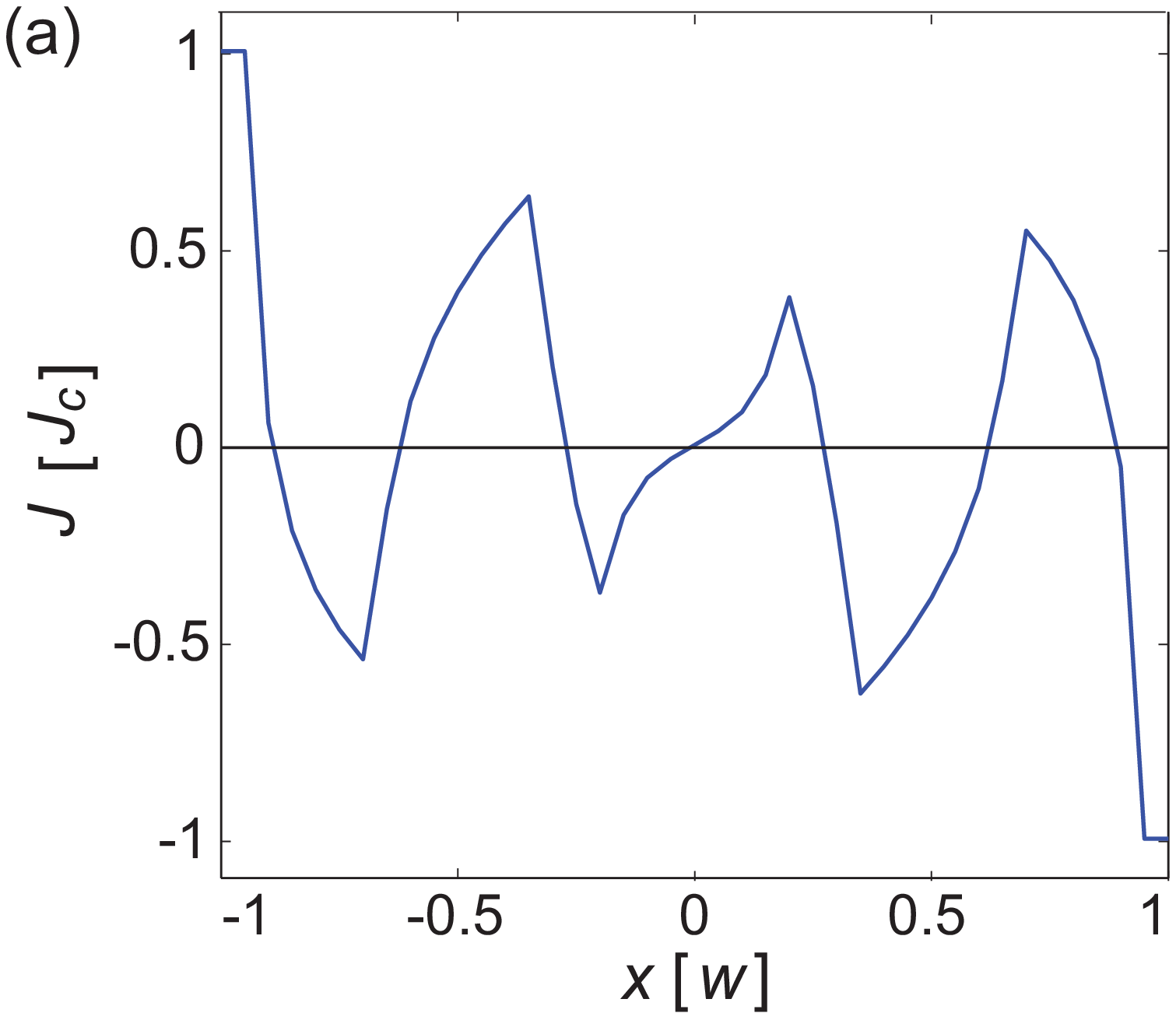}
\\ \vspace{0.2cm}
\includegraphics[width=5.6cm]{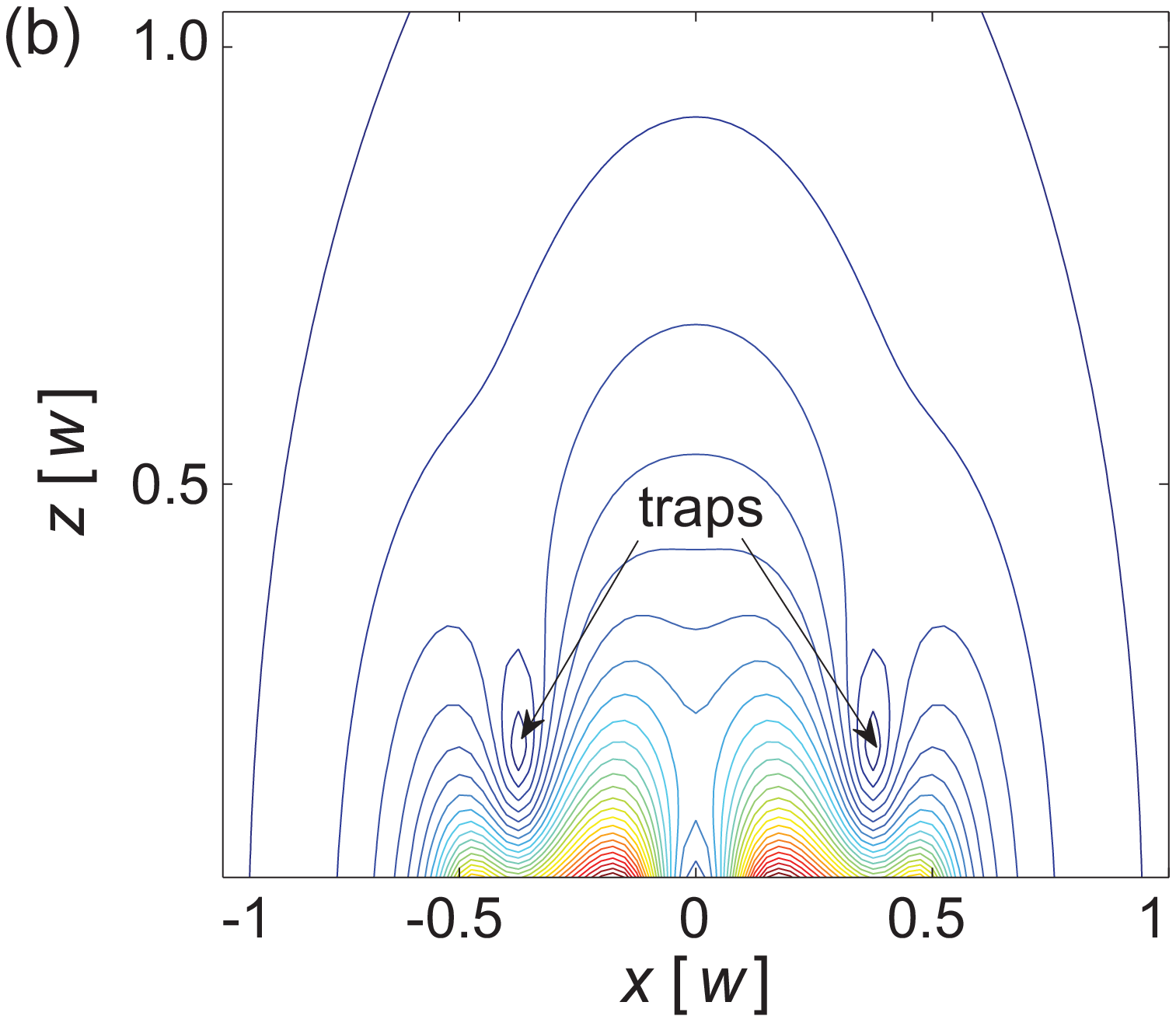}
\end{center} 
\vspace{-6mm} 
\caption{(color online) Loading field perpendicular to the strip with sequence $B_{ext}= 0\to 2.4B_c \to -1.2B_c \to 0.7B_c \to 0$. (a)  current density distribution in the strip. (b) equipotential lines for an atom with nonzero magnetic moment. A double trap can be seen around a height of $0.25w$. \label{fig:Bvortextrap2} }
\end{figure}

A double minimum in the potential can be realized with an additional external field reversal of $0.7B_c$ added to the presented sequence, as shown in Fig.~\ref{fig:Bvortextrap2}(b). This sequence consists in total of the fields $B_{ext}= 0\to 2.4B_c \to -1.2B_c \to 0.7B_c \to 0$, which leads to two additional segments with reversed current in the outer parts of the strip. The supercurrent density distribution is plotted in Fig.~\ref{fig:Bvortextrap2}(a). The trapping height is around $0.25w \sim 50 \; \mu$m.  These self-sufficient traps are attractive for integration with other atom-optic devices, due to the absence of noise associated with transport current. 

\section{Magnetization by transport current}
\label{sec:I}

\subsection{Induced supercurrent distribution and field distribution}
\label{sec:icurrent}

A transport current $I_{a}$ will also induce supercurrents  in the type-II superconductor. This can be used to generate various patterns of supercurrent distribution and corresponding magnetic potential surfaces. 

A transport current $I_{a}$ applied through a superconducting strip in the virgin state leads to magnetic flux penetrating the strip from the edges. In the region of flux penetration, the current density takes on the critical value $j_c$. In our case, the $z$-dependence of the current density is negligible as the thickness of the strip is much smaller than its width. The sheet current density $J(x)=j(x) d$ distribution in the superconducting strip is given by~\cite{Brandt} 
\begin{equation}
\label{eq:Brandt-I}
J(x,I_{a}) = \left \{ \begin{array}{ll} 
\frac{2J_{c}}{\pi}\arctan \left (\sqrt{\frac{w^2-b^2}{b^2-x^2}} \right ) , \quad & |x| \le b \\
J_{c} \; , & b \le |x| \le w
\end{array} \right.
\end{equation} 
where $b= w \sqrt{1-I_a^2/I_{p}^2}$ is the flux-free region, $I_a$ is the applied current and $I_{p} = 2wJ_{c}$ the maximal transport current. 
When the transport current $I_{a}$ decreases to $-I_{p}<I_{f}<I_{p}$, the total current density distribution can be described by~\cite{Brandt},
\begin{equation}
 J(x, I_{f}, J_c)=J(x,I_{a}, J_c)-J(x,I_{a}-I_{f}, 2J_c) \; . \label{eq:JdownI}
 \end{equation} 

\begin{figure}[h]
\begin{center}
\hspace{0.4cm}
\includegraphics[width=6.5cm]{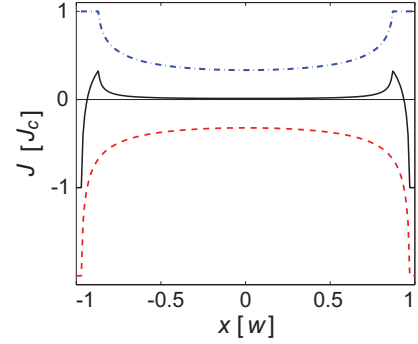}
\end{center} 
\vspace{-0.6cm}
\caption{(color online) Distribution of the sheet current density across the strip for the loading current pulse $I_{a}=0\to 0.5I_{p} \to 0$. Blue dot-dashed line: sheet current density for $I_a= 0  \rightarrow 0.5 I_p$; red dashed line: sheet current density for $I_a= 0.5 I_p  \rightarrow 0$; black solid line: finale sheet current density. The sheet current density is normalized by its critical value $J_c$ and the $x$-position is normalized by $w$. \label{fig:current-I} }
 \end{figure}

If the transport current is decreased to zero, some anti-vortices penetrate the strip from the edges and current density $-J(x,I_{a},2J_c)$ is induced in the opposite direction. We plot the corresponding remanent supercurrent density distribution in Fig.~\ref{fig:current-I}. In the outer region given by $w>|x|> 0.935w$, the sheet current density $J(x)$ is positive, whereas in the inner region given by $|x|< 0.935w$, $J(x)$ is negative. As the transport current $I_a$ is removed, the sum of the sheet current density across the strip width vanishes. The current flowing in the outer region $w>x> 0.935w$ ($-w<x<-0.935w$) has the same magnitude as the current flowing in the inner region $0<x< 0.935w$ ($-0.935w<x<0$). Therefore, the whole superconducting strip resembles four normally-conducting wires, where the outer two wires carry current in the opposite direction of the inner two wires. This schematic description of the current distribution is significantly different from the supercurrents induced by a perpendicular magnetic field pulse.

\begin{figure}[h]
\begin{center}
\includegraphics[width=6.1cm]{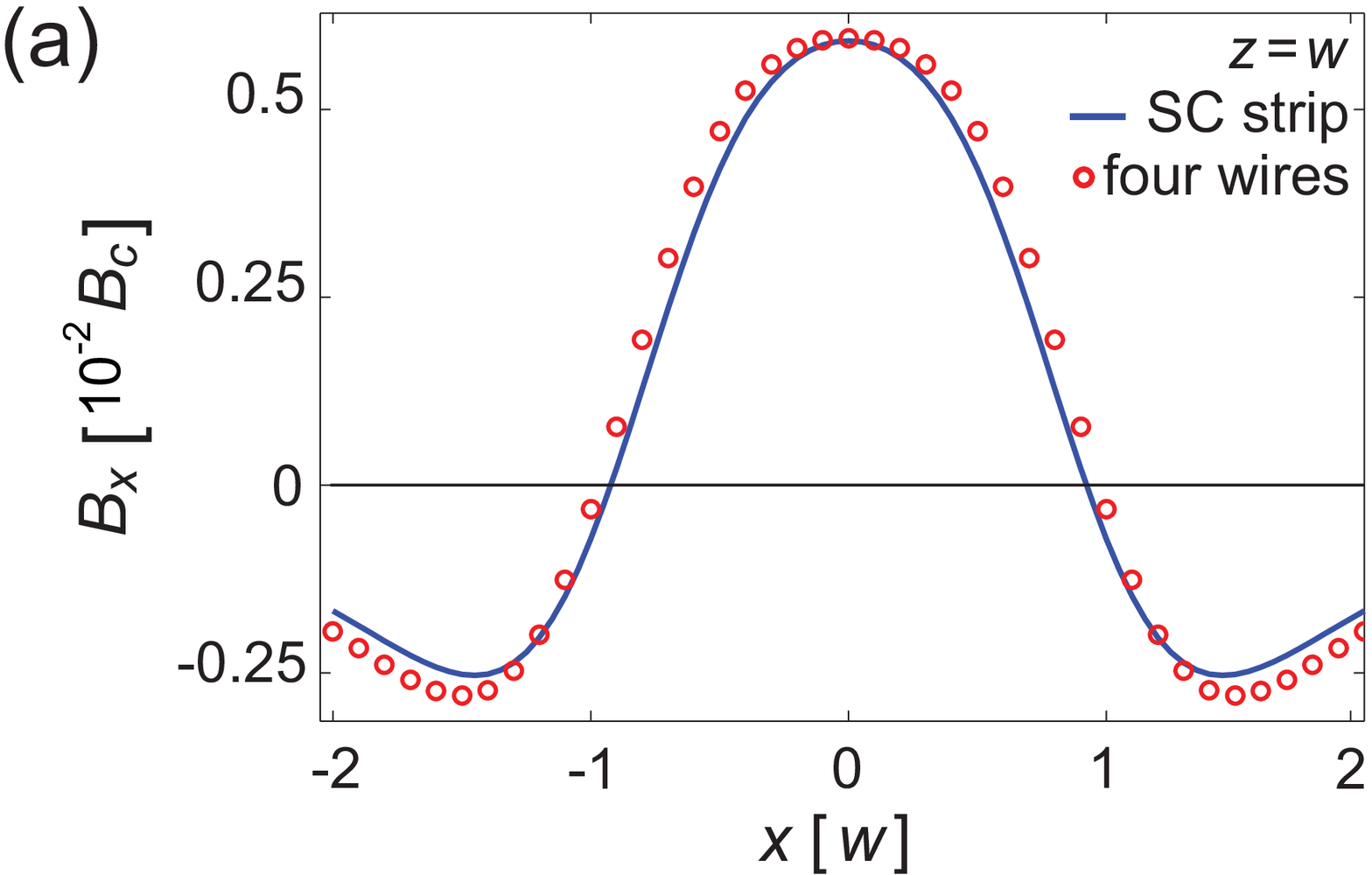}\\
\vspace{0.2cm}
\includegraphics[width=6.1cm]{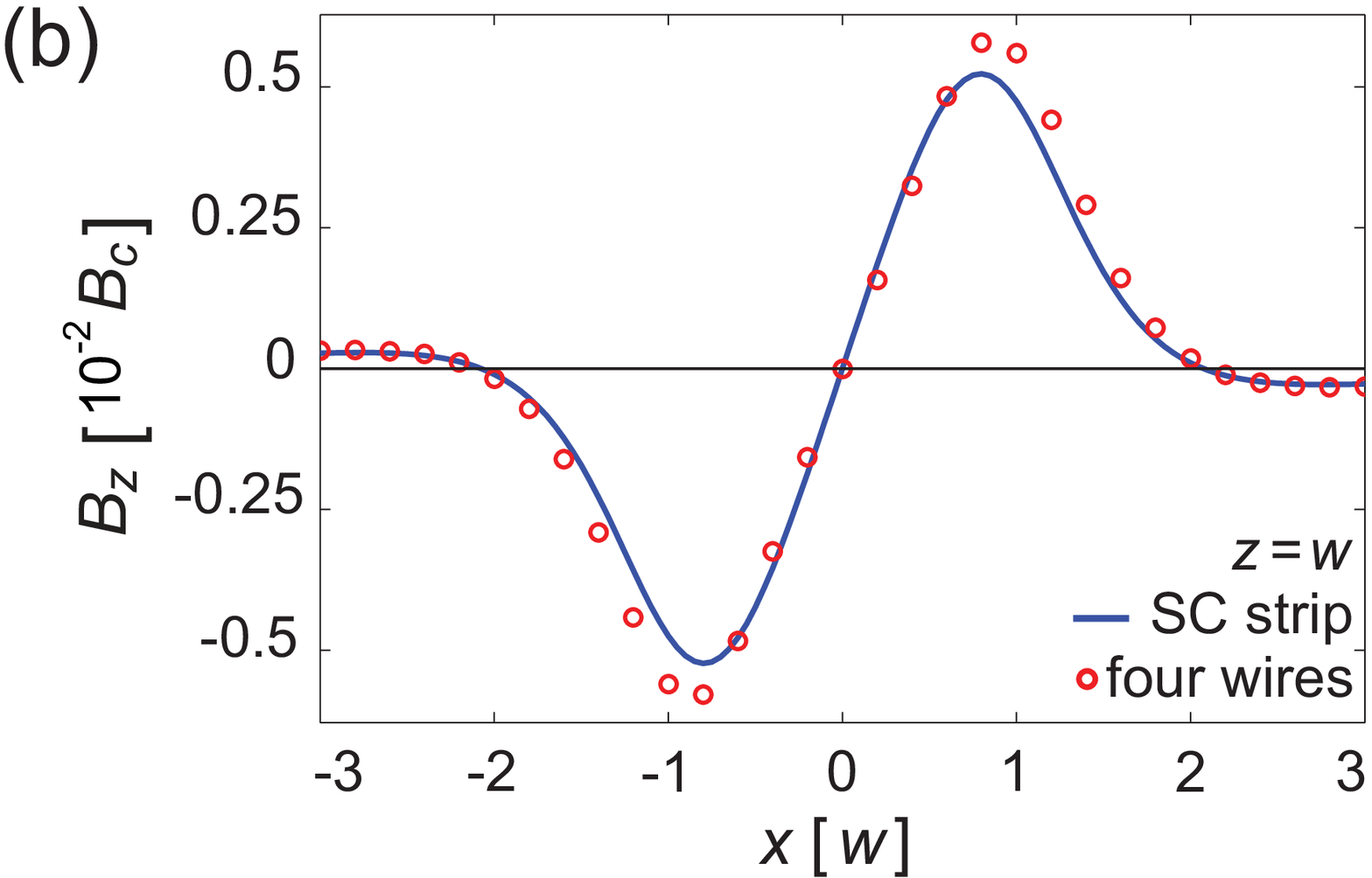}
\\ \vspace{0.2cm}
\includegraphics[width=6.1cm]{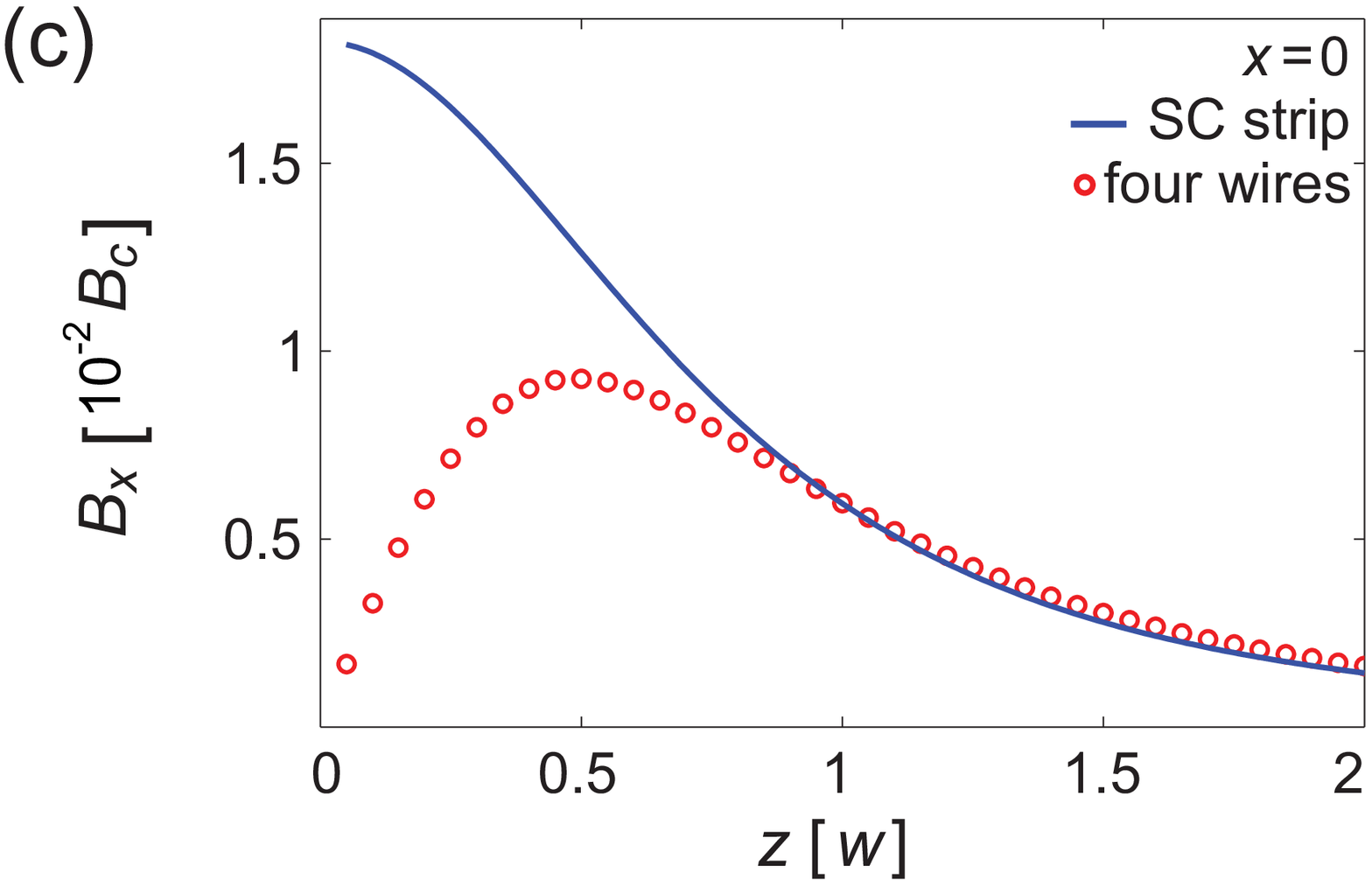}
\end{center} 
\vspace{-0.5cm}
\caption{(color online) Magnetic field components due to trapped vortices (blue lines) and due to four current-carrying wires (red circles). (a) $B_x(x,z)$ vs $x$ at fixed height $z=w$; (b) $B_z(x,z)$ vs $x$ at fixed height $z=w$; (c) $B_x(x,z)$ vs $z$ above the center of strip $x=0$. The applied current pulse is $I_{a}=0\to 0.5I_{p} \to 0$.  The four normally-conducting wires carry a current of $ \pm 0.04j_c d w$. The $x$- and $z$-positions are normalized by $w$ and the magnetic field is normalized by $\pi B_c$.  \label{fig:vortex-I}}
 \end{figure}

The magnetic field components generated by the supercurrents induced with a transport current are shown in Fig.~\ref{fig:vortex-I} (blue lines). For comparison, we also plot the magnetic field components generated by a setup of four wires in Fig.~\ref{fig:vortex-I} (red circles), where the current in each wire is equal to the current flowing in the region $-w<x<-0.935w$ of the superconducting strip. Integrating the sheet current density of Eq.~(\ref{eq:JdownI}) over $-w<x<-0.935w$ gives $I=0.04j_cdw$. We place the two outer wires at $\pm w$ and the two inner wires at $\pm 0.737w$. These positions yield good agreement as shown in Fig.~\ref{fig:vortex-I}.
%
%
 
 \subsection{Magnetic trap with bias field}
\label{sec:Itrap}
 
\begin{figure}[t]
\begin{center} 
\includegraphics[width=5.6cm]{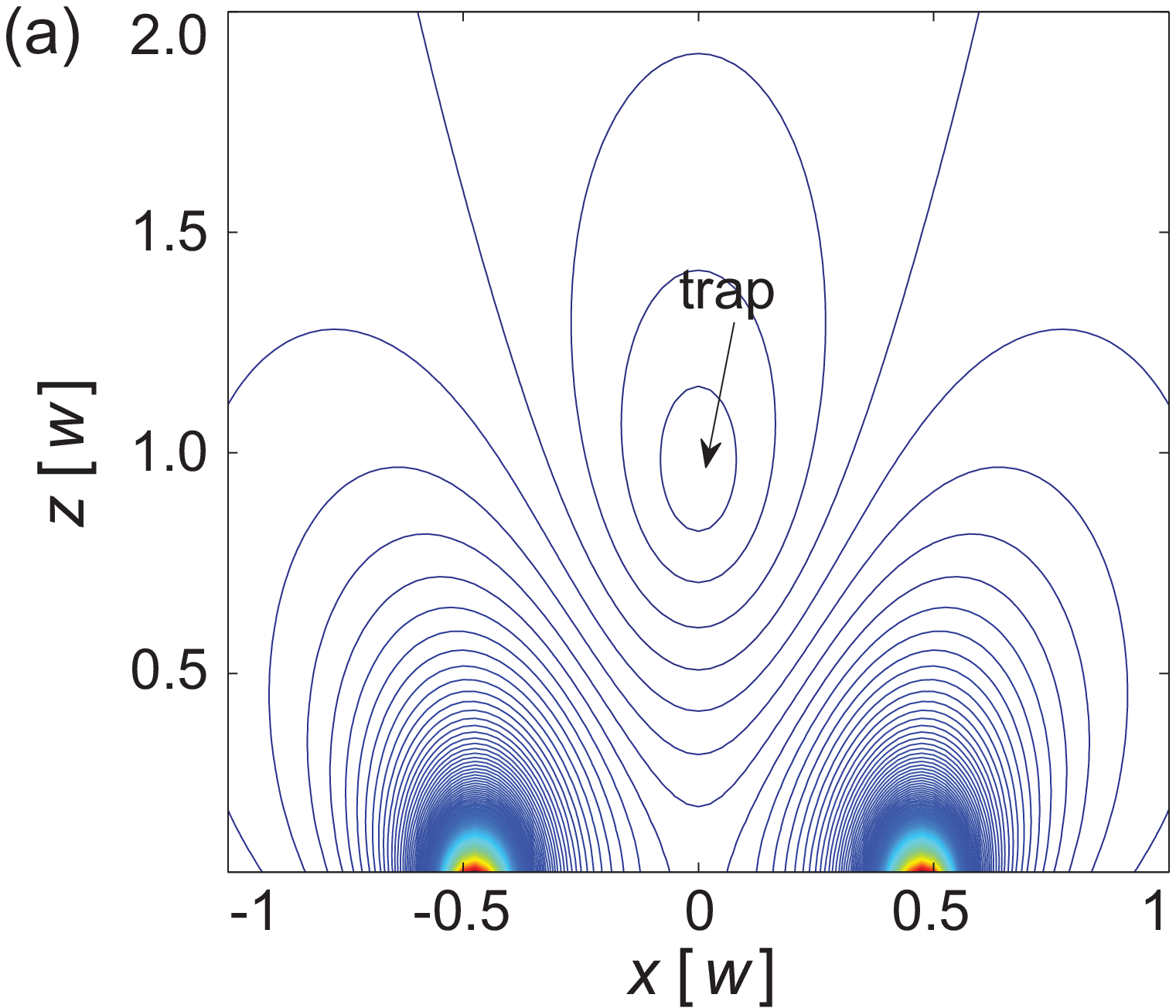}\\
\vspace{0.2cm}
\includegraphics[width=5.6cm]{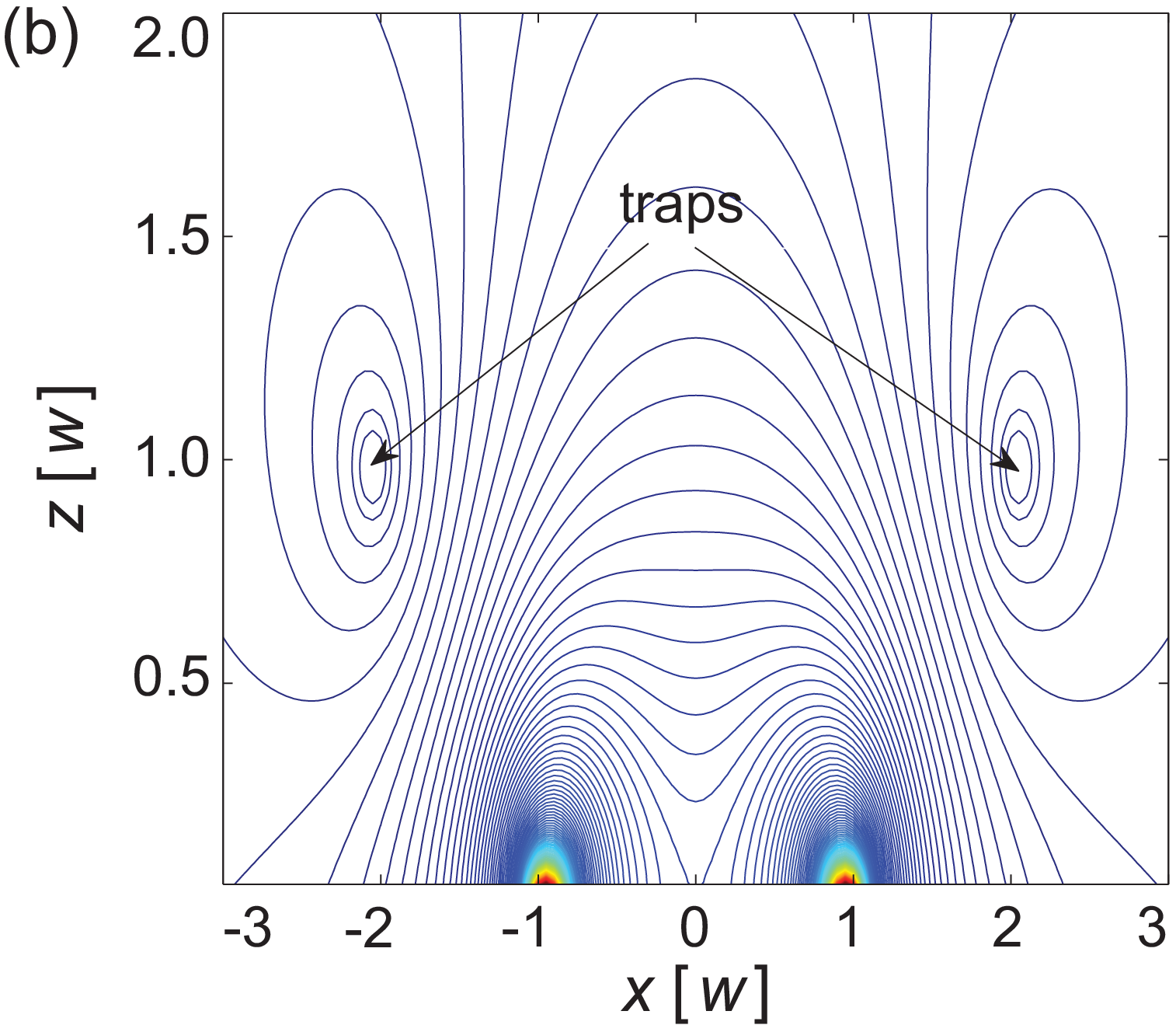}
\end{center}
\vspace{-6mm}
\caption{(color online) Equipotential lines of the total external field. (a) A single trap above the strip center formed by the vortex fields and a parallel field $B^x_{bias}=-0.006B_c$. (b) Double traps above $x= \pm 2w$ formed by the vortex fields, together with a parallel field $B^x_{bias}=0.0016B_c$. \label{fig:trap-I}  }
 \end{figure}

Trapping of atoms with the transport-current induced supercurrents can be achieved by applying a bias field $B_{bias}^x$ along the $x$-direction. This bias field cancels the non-zero field component $B_x(0,z)$ above the center of the strip, whereas the field component $B_z(0,z)$ always vanishes at the center, as shown in Figs.~\ref{fig:vortex-I}(a)-(b). Changing $B_{bias}^x$ cancels $B_x(0,z)$ at a different $z$, causing the trap to form at a different height. We plot $B_x(0,z)$ at different heights $z$ in Fig.~\ref{fig:vortex-I}(c). Due to the field repulsion caused by the Meissner effect, this component is qualitatively different from the potential of four normally-conducting current-carrying wires. A typical example of the newly designed trap, formed by $B^x_{bias}= -0.006 B_c$ at $z=w$ is shown in Fig.~\ref{fig:trap-I}(a). Simulations with our superconducting sample strip yield a loading transport current pulse $I_a=I_p=8.32$~A and a bias field $B_{bias}^x=-0.5$~G generating a magnetic trap at $z=200 \; \mu$m above the strip center. 
 
A double magnetic trap can be realized by applying a bias field~$B^x_{bias}$. The $z$-component of the magnetic field generated by the induced supercurrents also crosses zero around $\pm 2.0w$, as shown Fig.~\ref{fig:vortex-I}(b), while Fig.~\ref{fig:vortex-I}(b) shows that the $x$-component $B_x(\pm 2.0w,w)$ has opposite sign at those points. The application of a bias field $B^x_{bias}=-B_x(\pm 2.0w,w)= 0.0016 B_c$ generates a double minimum in the potential at $(\pm 2.0w, w)$ as presented in Fig.~\ref{fig:trap-I}(b). 
Simulations with our superconducting sample strip yield a loading transport current pulse $I_a=I_p=8.32$~A and a bias field $B_{bias}^x=0.13$~G generating a magnetic trap at a distance $z=200 \; \mu$m and lateral position $x=\pm 400 \; \mu$m.

As the sum of induced supercurrents across each half of the strip width vanishes, 
the magnetic fields generated by opposing supercurrents cancel each other to much greater extent when compared to supercurrents induced by magnetic fields. The gradient fields required to form the confining potentials arise only from the spatial distribution of the induced supercurrents. Therefore, the confinement is much weaker now with respect to the trapping potentials discussed in Sec.~\ref{sec:B}. Efficient trapping with transport-current induced vortex potentials can in principle be achieved with sufficiently cold atoms. This confinement can also be enhanced with superconductors offering higher critical currents.
 
\section{Effects of transport currents on magnetic trap properties}
\label{sec:discuss}

\subsection{Influence of hysteresis on the trap properties using a transport current}
\label{sec:hys}
Recent experiments on superconducting atom chips have demonstrated magnetic traps created by a superconducting wire carrying a transport current and a homogenous bias field~\cite{Emmert09,Hufnagel,Kasch}. In these experiments the applied magnetic fields and transport currents are typically varied during the course of loading, transferring, and trapping cold atoms. Due to the memory effect of the superconductor in the mixed state, the influence of the induced supercurrents on the total trapping potential has to be considered~\cite{Dikovsky,Brandt} and has recently been investigated experimentally~\cite{Emmert09}.

\begin{figure}[h]
\begin{center}
\includegraphics[width=6.5cm]{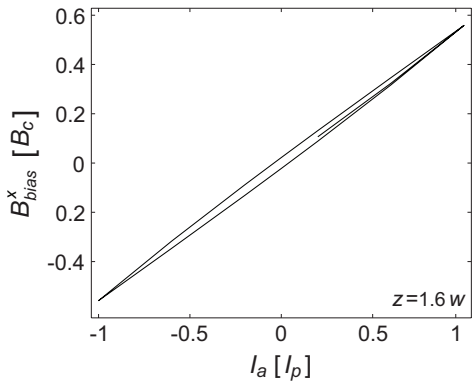}
\end{center}
\vspace{-0.6cm}
\caption{Bias field required to form a trap at fixed height $1.6w$ applying various transport current histories. The transport current increases from zero to the maximum value $I_{p}$, then decreases to $-I_{p}$, before again returning to $I_{p}$. \label{fig:hys} }
 \end{figure} 
Here we analyze the influence of the memory effect of variable transport currents on the micro-trap, that is generated by combining a transport current with a homogenous bias field parallel to its surface as in ~\cite{Emmert09,Kasch}. Our studies show that the memory effect of the transport currents has a significant effect on the total trapping potential. In our simulation we cycle the transport current $I_a$ from zero to the maximum current $I_{p}$, then reverse it to $-I_{p}$ and increase it back to $I_{p}$. The induced supercurrent and the overall magnetic potential generated by the wire changes with different applied current histories. To visualize changes in the total trapping potential caused by the hysteresis, we calculate the bias field required to form a trap at a fixed height for different current histories. 
In Fig.~\ref{fig:hys}, we plot the bias field for a fixed height of $z=1.6w$. For our sample strip, the maximum applied current is $I_p=8.32$~A. Increasing the transport current from zero to $I_a=0.2I_p=1.66$~A, a bias field of $B_{bias}^x=0.106B_c=8.84$~G is required to form a field minimum at a height of $z=1.6w=320 \; \mu$m. Increasing the transport current to $I_a=I_p=8.32$~A and then decreasing it back to $I_a=0.2I_p=1.66$~A, a higher bias field of $B_{bias}^x=0.13B_c=10.94$~G is required at the same trap-to-surface distance. Further decreasing the transport current to $I_a=-I_p=-8.32$~A and then increasing it to $I_a=0.2I_p=1.66$~A again, a lower bias field of $B_{bias}^x=0.089B_c=7.38$~G is needed. Note that when the transport current is around zero, the superconducting strip is in the remanent state and the trapping field is provided by the supercurrents in the strip as discussed in Sec.~\ref{sec:Itrap}. However, for small transport currents the confinement can be so weak that atoms escape from the trap. The memory effect of the strip and its effect on the total trapping potential height is clearly evident in the simulations shown in Fig.~\ref{fig:hys}.

\subsection{Effect of transport current on trapped vortices}
Transport currents can also trigger motion and dissipation of trapped vortices. These effects are more complicated, as they also depend on the depinning of vortices, which in turn depends on the pinning energy and defects. In our model these effects are not considered and here we simply present a few experimental measurements. We experimentally observe the influence of transport current on remanent magnetization in the same magnetic trap type as described in Sec.~\ref{sec:Bpara}. 
\begin{figure}[h]
\begin{center} 
\includegraphics[width=6cm]{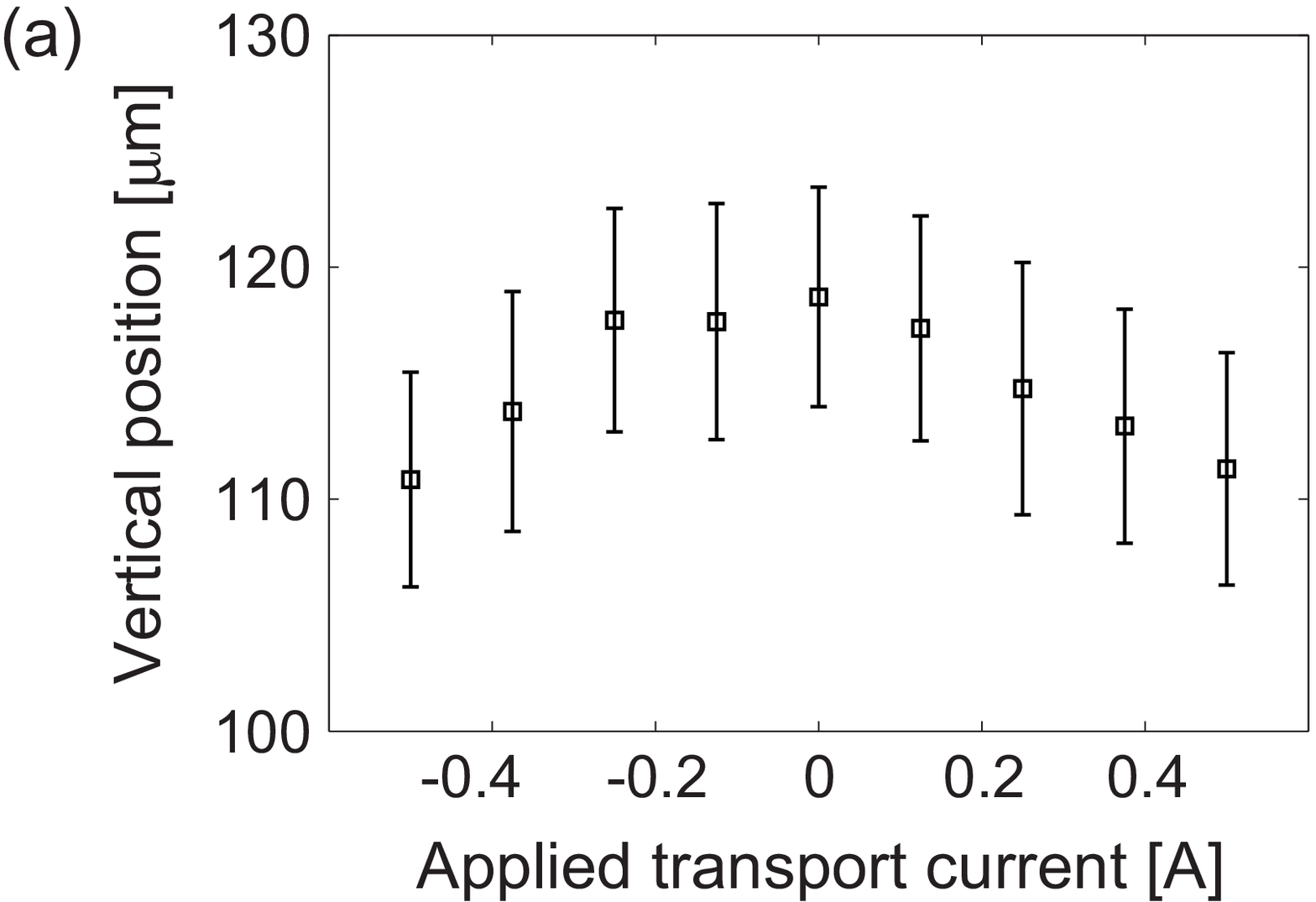}\\
\vspace{0.2cm}
\includegraphics[width=6cm]{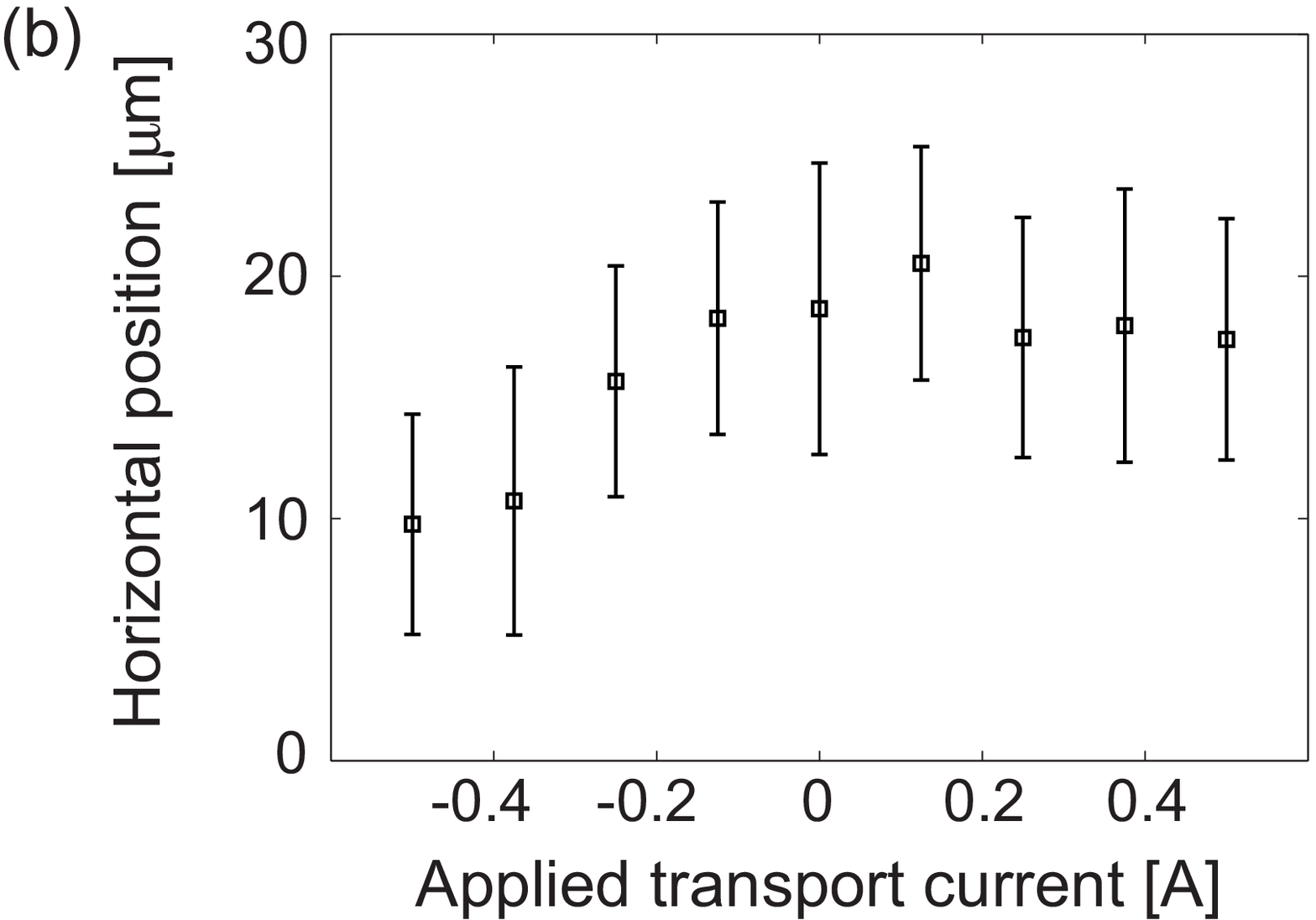}
\end{center}   
\vspace{-0.7cm}\caption{\label{fig:5} effect of supercurrents due to pulsed transport current visualized by the position of the micro-trap resembling the Z-wire geometry. The trap-to-surface distance is measured absolute with a reflection method from the chip surface while the horizontal position is measured with an arbitrary offset. Error bars reflect systematic and statistical uncertainty.}
 \end{figure}
First we trap ultra-cold atoms in this trap for $100$~ms relying only on induced supercurrents. Then we apply a variable transport current through the strip for a duration of $200$~ms. Afterward the transport current is switched off and the atoms are held $100$~ms longer in the trap relying only on the remaining induced supercurrents. This procedure ensures that the trapping potential is no longer directly influenced by the transport current. However, the changes of the induced supercurrents due to the applied transport current remain and are reflected in the trapping potential for the atoms.

We visualize these changes by a measurement of the trap position, shown in Fig.~\ref{fig:5}. We apply a magnetic field pulse of $+220$~G before each realization of the micro-trap, ensuring identical vortex density before each measurement. We then apply a transport current variable from $0.5$~A to $-0.5$~A. As shown in Fig.~\ref{fig:5}(a), the trap-to-surface distance is reduced for higher absolute value of the applied current, independent of the current direction. The reduced distance is caused by a decrease in magnitude of the induced supercurrents. The cloud is also displaced horizontally as shown in Fig.~\ref{fig:5}(b). However, the horizontal shift is only detectable for negative currents within our measurement uncertainty. We infer that the transport current not only reduces the remanent magnetization, but also produces spatial redistribution in the film. This shows that the transport current can trigger vortex annihilation as well as more complex dynamics.\\

\section{Conclusion}
\label{sec:con}

In conclusion, we have investigated the distribution of supercurrents and vortices in a superconducting strip for the purpose of designing magnetic trapping potentials for neutral atoms.  The supercurrents can be induced by applying an external magnetic field perpendicularly to the strip surface or by directly applying a transport current. Various supercurrent distributions can be induced in the strip by different vortex-loading procedures, all due to the memory effect of the superconductor. This versatility may be exploited to generate different spatial magnetic field patterns, with the goal of tailoring magnetic trapping potentials for atoms.  The properties of the vortex traps, such as trap position and the trapping geometry are strongly dependent on the detailed distribution of vortices.  Cold atoms may be used in future work as a probe of the detailed vortex distribution.  A new kind of self-sufficient atom trap has been designed with this technique. One major advantage of the vortex trap is that the absence of transport currents and external bias fields may reduce or even eliminate the technical noise due to unstable power sources. For this reason, these self-sufficient traps are attractive candidates for integration with other devices for cold atom manipulation. 

\begin{acknowledgments}
We acknowledge financial support from Nanyang Technological University (grant no. WBS M58110036), A-Star (grant no. SERC 072 101 0035 and WBS R-144-000-189-305) and the Centre for Quantum Technologies, Singapore. MJL acknowledges travel support from NSF PHY-0613659 and the Rowan University NSFG program. 
\end{acknowledgments}



\begin{thebibliography}{29}
\expandafter\ifx\csname natexlab\endcsname\relax\def\natexlab#1{#1}\fi
\expandafter\ifx\csname bibnamefont\endcsname\relax
  \def\bibnamefont#1{#1}\fi
\expandafter\ifx\csname bibfnamefont\endcsname\relax
  \def\bibfnamefont#1{#1}\fi
\expandafter\ifx\csname citenamefont\endcsname\relax
  \def\citenamefont#1{#1}\fi
\expandafter\ifx\csname url\endcsname\relax
  \def\url#1{\texttt{#1}}\fi
\expandafter\ifx\csname urlprefix\endcsname\relax\def\urlprefix{URL }\fi
\providecommand{\bibinfo}[2]{#2}
\providecommand{\eprint}[2][]{\url{#2}}

\bibitem[{\citenamefont{Hinds and Hughes}(1999)}]{Hinds}
\bibinfo{author}{\bibfnamefont{E.~A.} \bibnamefont{Hinds}} \bibnamefont{and}
  \bibinfo{author}{\bibfnamefont{I.~G.} \bibnamefont{Hughes}},
  \bibinfo{journal}{J.\ Phys.\ D\ Appl.\ Phys.} \textbf{\bibinfo{volume}{32}},
  \bibinfo{pages}{R119} (\bibinfo{year}{1999}).

\bibitem[{\citenamefont{Folman et~al.}(2002)\citenamefont{Folman, Kr{\"u}ger,
  Schmiedmayer, Denschlag, and Henkel}}]{Folman02}
\bibinfo{author}{\bibfnamefont{R.}~\bibnamefont{Folman}},
  \bibinfo{author}{\bibfnamefont{P.}~\bibnamefont{Kr{\"u}ger}},
  \bibinfo{author}{\bibfnamefont{J.}~\bibnamefont{Schmiedmayer}},
  \bibinfo{author}{\bibfnamefont{J.}~\bibnamefont{Denschlag}},
  \bibnamefont{and} \bibinfo{author}{\bibfnamefont{C.}~\bibnamefont{Henkel}},
  \bibinfo{journal}{Adv.\ At.\ Mol.\ Opt.\ Phys.}
  \textbf{\bibinfo{volume}{48}}, \bibinfo{pages}{263} (\bibinfo{year}{2002}).

\bibitem[{\citenamefont{Fort{\'a}gh and Zimmermann}(2007)}]{Fortagh07}
\bibinfo{author}{\bibfnamefont{J.}~\bibnamefont{Fort{\'a}gh}} \bibnamefont{and}
  \bibinfo{author}{\bibfnamefont{C.}~\bibnamefont{Zimmermann}},
  \bibinfo{journal}{Rev.\ Mod.\ Phys.} \textbf{\bibinfo{volume}{79}},
  \bibinfo{pages}{235} (\bibinfo{year}{2007}).

\bibitem[{\citenamefont{Henkel et~al.}(1999)\citenamefont{Henkel, P{\"o}tting,
  and Wilkens}}]{Henkel99c}
\bibinfo{author}{\bibfnamefont{C.}~\bibnamefont{Henkel}},
  \bibinfo{author}{\bibfnamefont{S.}~\bibnamefont{P{\"o}tting}},
  \bibnamefont{and} \bibinfo{author}{\bibfnamefont{M.}~\bibnamefont{Wilkens}},
  \bibinfo{journal}{Appl. Phys. B} \textbf{\bibinfo{volume}{69}},
  \bibinfo{pages}{379} (\bibinfo{year}{1999}).

\bibitem[{\citenamefont{Jones et~al.}(2003)\citenamefont{Jones, Vale, Sahagun,
  Hall, and Hinds}}]{Jones}
\bibinfo{author}{\bibfnamefont{M.~P.~A.} \bibnamefont{Jones}},
  \bibinfo{author}{\bibfnamefont{C.~J.} \bibnamefont{Vale}},
  \bibinfo{author}{\bibfnamefont{D.}~\bibnamefont{Sahagun}},
  \bibinfo{author}{\bibfnamefont{B.~V.} \bibnamefont{Hall}}, \bibnamefont{and}
  \bibinfo{author}{\bibfnamefont{E.~A.} \bibnamefont{Hinds}},
  \bibinfo{journal}{Phys. Rev. Lett.} \textbf{\bibinfo{volume}{91}},
  \bibinfo{pages}{080401} (\bibinfo{year}{2003}).

\bibitem[{\citenamefont{Harber et~al.}(2003)\citenamefont{Harber, McGuirk,
  Obrecht, and Cornell}}]{Harber}
\bibinfo{author}{\bibfnamefont{D.~M.} \bibnamefont{Harber}},
  \bibinfo{author}{\bibfnamefont{J.~M.} \bibnamefont{McGuirk}},
  \bibinfo{author}{\bibfnamefont{J.~M.} \bibnamefont{Obrecht}},
  \bibnamefont{and} \bibinfo{author}{\bibfnamefont{E.~A.}
  \bibnamefont{Cornell}}, \bibinfo{journal}{J. Low Temp. Phys.}
  \textbf{\bibinfo{volume}{133}}, \bibinfo{pages}{229} (\bibinfo{year}{2003}).

\bibitem[{\citenamefont{Skagerstam et~al.}(2006)\citenamefont{Skagerstam,
  Hohenester, Eiguren, and Rekdal}}]{Skagerstam06a}
\bibinfo{author}{\bibfnamefont{B.-S.~K.} \bibnamefont{Skagerstam}},
  \bibinfo{author}{\bibfnamefont{U.}~\bibnamefont{Hohenester}},
  \bibinfo{author}{\bibfnamefont{A.}~\bibnamefont{Eiguren}}, \bibnamefont{and}
  \bibinfo{author}{\bibfnamefont{P.~K.} \bibnamefont{Rekdal}},
  \bibinfo{journal}{Phys.\ Rev.\ Lett.} \textbf{\bibinfo{volume}{97}},
  \bibinfo{pages}{070401} (\bibinfo{year}{2006}).

\bibitem[{\citenamefont{Hohenester et~al.}(2007)\citenamefont{Hohenester,
  Eiguren, Scheel, and Hinds}}]{Hohenester07}
\bibinfo{author}{\bibfnamefont{U.}~\bibnamefont{Hohenester}},
  \bibinfo{author}{\bibfnamefont{A.}~\bibnamefont{Eiguren}},
  \bibinfo{author}{\bibfnamefont{S.}~\bibnamefont{Scheel}}, \bibnamefont{and}
  \bibinfo{author}{\bibfnamefont{E.~A.} \bibnamefont{Hinds}},
  \bibinfo{journal}{Phys. Rev. A} \textbf{\bibinfo{volume}{76}}
  (\bibinfo{year}{2007}).

\bibitem[{\citenamefont{Mukai et~al.}(2007)\citenamefont{Mukai, Hufnagel,
  Kasper, Meno, Tsukada, Semba, and Shimizu}}]{Mukai}
\bibinfo{author}{\bibfnamefont{T.}~\bibnamefont{Mukai}},
  \bibinfo{author}{\bibfnamefont{C.}~\bibnamefont{Hufnagel}},
  \bibinfo{author}{\bibfnamefont{A.}~\bibnamefont{Kasper}},
  \bibinfo{author}{\bibfnamefont{T.}~\bibnamefont{Meno}},
  \bibinfo{author}{\bibfnamefont{A.}~\bibnamefont{Tsukada}},
  \bibinfo{author}{\bibfnamefont{K.}~\bibnamefont{Semba}}, \bibnamefont{and}
  \bibinfo{author}{\bibfnamefont{F.}~\bibnamefont{Shimizu}},
  \bibinfo{journal}{Phys.\ Rev.\ Lett.} \textbf{\bibinfo{volume}{98}},
  \bibinfo{pages}{260407} (\bibinfo{year}{2007}).

\bibitem[{\citenamefont{Nirrengarten et~al.}(2006)\citenamefont{Nirrengarten,
  Qarry, Roux, Emmert, Nogues, Brune, {J.-M. Raimond}, and
  Haroche}}]{Nirrengarten}
\bibinfo{author}{\bibfnamefont{T.}~\bibnamefont{Nirrengarten}},
  \bibinfo{author}{\bibfnamefont{A.}~\bibnamefont{Qarry}},
  \bibinfo{author}{\bibfnamefont{C.}~\bibnamefont{Roux}},
  \bibinfo{author}{\bibfnamefont{A.}~\bibnamefont{Emmert}},
  \bibinfo{author}{\bibfnamefont{G.}~\bibnamefont{Nogues}},
  \bibinfo{author}{\bibfnamefont{M.}~\bibnamefont{Brune}},
  \bibinfo{author}{\bibnamefont{{J.-M. Raimond}}}, \bibnamefont{and}
  \bibinfo{author}{\bibfnamefont{S.}~\bibnamefont{Haroche}},
  \bibinfo{journal}{Phys.\ Rev.\ Lett.} \textbf{\bibinfo{volume}{97}},
  \bibinfo{pages}{200405} (\bibinfo{year}{2006}).

\bibitem[{\citenamefont{Roux et~al.}(2008)\citenamefont{Roux, Emmert, Lupa{\c
  s}cu, Nirrengarten, Nogues, Brune, {J.-M. Raimond}, and Haroche}}]{Roux}
\bibinfo{author}{\bibfnamefont{C.}~\bibnamefont{Roux}},
  \bibinfo{author}{\bibfnamefont{A.}~\bibnamefont{Emmert}},
  \bibinfo{author}{\bibfnamefont{A.}~\bibnamefont{Lupa{\c s}cu}},
  \bibinfo{author}{\bibfnamefont{T.}~\bibnamefont{Nirrengarten}},
  \bibinfo{author}{\bibfnamefont{G.}~\bibnamefont{Nogues}},
  \bibinfo{author}{\bibfnamefont{M.}~\bibnamefont{Brune}},
  \bibinfo{author}{\bibnamefont{{J.-M. Raimond}}}, \bibnamefont{and}
  \bibinfo{author}{\bibfnamefont{S.}~\bibnamefont{Haroche}},
  \bibinfo{journal}{Eur.\ Phys.\ Lett.} \textbf{\bibinfo{volume}{81}},
  \bibinfo{pages}{56004} (\bibinfo{year}{2008}).

\bibitem[{\citenamefont{Kasch et~al.}(2009)\citenamefont{Kasch, Hattermann,
  Cano, Judd, Scheel, Zimmermann, Kleiner, K{\"o}lle, and Fort{\'a}gh}}]{Kasch}
\bibinfo{author}{\bibfnamefont{B.}~\bibnamefont{Kasch}},
  \bibinfo{author}{\bibfnamefont{H.}~\bibnamefont{Hattermann}},
  \bibinfo{author}{\bibfnamefont{D.}~\bibnamefont{Cano}},
  \bibinfo{author}{\bibfnamefont{T.~E.} \bibnamefont{Judd}},
  \bibinfo{author}{\bibfnamefont{S.}~\bibnamefont{Scheel}},
  \bibinfo{author}{\bibfnamefont{C.}~\bibnamefont{Zimmermann}},
  \bibinfo{author}{\bibfnamefont{R.}~\bibnamefont{Kleiner}},
  \bibinfo{author}{\bibfnamefont{D.}~\bibnamefont{K{\"o}lle}},
  \bibnamefont{and}
  \bibinfo{author}{\bibfnamefont{J.}~\bibnamefont{Fort{\'a}gh}}
  (\bibinfo{year}{2009}), \eprint{arXiv:0906.1369}.

\bibitem[{\citenamefont{Fermani et~al.}(2010)\citenamefont{Fermani, {M\"u}ller,
  Zhang, Lim, and Dumke}}]{Fermani09}
\bibinfo{author}{\bibfnamefont{R.}~\bibnamefont{Fermani}},
  \bibinfo{author}{\bibfnamefont{T.}~\bibnamefont{{M\"u}ller}},
  \bibinfo{author}{\bibfnamefont{B.}~\bibnamefont{Zhang}},
  \bibinfo{author}{\bibfnamefont{M.~J.} \bibnamefont{Lim}}, \bibnamefont{and}
  \bibinfo{author}{\bibfnamefont{R.}~\bibnamefont{Dumke}},
  \bibinfo{journal}{J.\ Phys. \ B}  (\bibinfo{year}{2010}),
  \eprint{arXiv:0912.2183}.

\bibitem[{\citenamefont{Scheel et~al.}(2005)\citenamefont{Scheel, Rekdal,
  Knight, and Hinds}}]{Scheel05a}
\bibinfo{author}{\bibfnamefont{S.}~\bibnamefont{Scheel}},
  \bibinfo{author}{\bibfnamefont{P.~K.} \bibnamefont{Rekdal}},
  \bibinfo{author}{\bibfnamefont{P.~L.} \bibnamefont{Knight}},
  \bibnamefont{and} \bibinfo{author}{\bibfnamefont{E.~A.} \bibnamefont{Hinds}},
  \bibinfo{journal}{Phys. Rev. A} \textbf{\bibinfo{volume}{72}},
  \bibinfo{pages}{042901} (\bibinfo{year}{2005}).

\bibitem[{\citenamefont{Scheel et~al.}(2007)\citenamefont{Scheel, Fermani, and
  Hinds}}]{Scheel07}
\bibinfo{author}{\bibfnamefont{S.}~\bibnamefont{Scheel}},
  \bibinfo{author}{\bibfnamefont{R.}~\bibnamefont{Fermani}}, \bibnamefont{and}
  \bibinfo{author}{\bibfnamefont{E.~A.} \bibnamefont{Hinds}},
  \bibinfo{journal}{Phys.\ Rev.\ A} \textbf{\bibinfo{volume}{75}},
  \bibinfo{pages}{064901} (\bibinfo{year}{2007}).

\bibitem[{\citenamefont{Tian et~al.}(2004)\citenamefont{Tian, Rabl, Blatt, and
  Zoller}}]{Tian}
\bibinfo{author}{\bibfnamefont{L.}~\bibnamefont{Tian}},
  \bibinfo{author}{\bibfnamefont{P.}~\bibnamefont{Rabl}},
  \bibinfo{author}{\bibfnamefont{R.}~\bibnamefont{Blatt}}, \bibnamefont{and}
  \bibinfo{author}{\bibfnamefont{P.}~\bibnamefont{Zoller}},
  \bibinfo{journal}{Phys.\ Rev.\ Lett.} \textbf{\bibinfo{volume}{92}},
  \bibinfo{pages}{247902} (\bibinfo{year}{2004}).

\bibitem[{\citenamefont{Andr{\'e} et~al.}(2006)\citenamefont{Andr{\'e},
  DeMille, Doyle, Lukin, Maxwell, P.Rabl, Schoelkopf, and Zoller}}]{Andre}
\bibinfo{author}{\bibfnamefont{A.}~\bibnamefont{Andr{\'e}}},
  \bibinfo{author}{\bibfnamefont{D.}~\bibnamefont{DeMille}},
  \bibinfo{author}{\bibfnamefont{J.~M.} \bibnamefont{Doyle}},
  \bibinfo{author}{\bibfnamefont{M.~D.} \bibnamefont{Lukin}},
  \bibinfo{author}{\bibfnamefont{S.~E.} \bibnamefont{Maxwell}},
  \bibinfo{author}{\bibnamefont{P.Rabl}}, \bibinfo{author}{\bibfnamefont{R.~J.}
  \bibnamefont{Schoelkopf}}, \bibnamefont{and}
  \bibinfo{author}{\bibfnamefont{P.}~\bibnamefont{Zoller}},
  \bibinfo{journal}{Nature\ Phys.} \textbf{\bibinfo{volume}{2}},
  \bibinfo{pages}{636} (\bibinfo{year}{2006}).

\bibitem[{\citenamefont{Tordrup and M{\o}lmer}(2008)}]{Tordrup}
\bibinfo{author}{\bibfnamefont{K.}~\bibnamefont{Tordrup}} \bibnamefont{and}
  \bibinfo{author}{\bibfnamefont{K.}~\bibnamefont{M{\o}lmer}},
  \bibinfo{journal}{Phys.\ Rev.\ A} \textbf{\bibinfo{volume}{77}},
  \bibinfo{pages}{020301(R)} (\bibinfo{year}{2008}).

\bibitem[{\citenamefont{Verd{\'u} et~al.}(2009)\citenamefont{Verd{\'u}, Zoubi,
  {Ch. Koller}, Majer, Ritsch, and Schmiedmayer}}]{Verdu}
\bibinfo{author}{\bibfnamefont{J.}~\bibnamefont{Verd{\'u}}},
  \bibinfo{author}{\bibfnamefont{H.}~\bibnamefont{Zoubi}},
  \bibinfo{author}{\bibnamefont{{Ch. Koller}}},
  \bibinfo{author}{\bibfnamefont{J.}~\bibnamefont{Majer}},
  \bibinfo{author}{\bibfnamefont{H.}~\bibnamefont{Ritsch}}, \bibnamefont{and}
  \bibinfo{author}{\bibfnamefont{J.}~\bibnamefont{Schmiedmayer}},
  \bibinfo{journal}{Phys.\ Rev.\ Lett.} \textbf{\bibinfo{volume}{103}},
  \bibinfo{pages}{043603} (\bibinfo{year}{2009}).

\bibitem[{\citenamefont{Dikovsky et~al.}(2009)\citenamefont{Dikovsky,
  Sokolovsky, Zhang, Henkel, and Folman}}]{Dikovsky}
\bibinfo{author}{\bibfnamefont{V.}~\bibnamefont{Dikovsky}},
  \bibinfo{author}{\bibfnamefont{V.}~\bibnamefont{Sokolovsky}},
  \bibinfo{author}{\bibfnamefont{B.}~\bibnamefont{Zhang}},
  \bibinfo{author}{\bibfnamefont{C.}~\bibnamefont{Henkel}}, \bibnamefont{and}
  \bibinfo{author}{\bibfnamefont{R.}~\bibnamefont{Folman}},
  \bibinfo{journal}{Eur.\ Phys.\ J.\ D} \textbf{\bibinfo{volume}{51}},
  \bibinfo{pages}{247} (\bibinfo{year}{2009}).

\bibitem[{\citenamefont{Emmert et~al.}(2009)\citenamefont{Emmert, Lupa{\c s}cu,
  Nogues, Brune, {J.-M. Raimond}, and Haroche}}]{Emmert09}
\bibinfo{author}{\bibfnamefont{A.}~\bibnamefont{Emmert}},
  \bibinfo{author}{\bibfnamefont{A.}~\bibnamefont{Lupa{\c s}cu}},
  \bibinfo{author}{\bibfnamefont{G.}~\bibnamefont{Nogues}},
  \bibinfo{author}{\bibfnamefont{M.}~\bibnamefont{Brune}},
  \bibinfo{author}{\bibnamefont{{J.-M. Raimond}}}, \bibnamefont{and}
  \bibinfo{author}{\bibfnamefont{S.}~\bibnamefont{Haroche}},
  \bibinfo{journal}{Eur.\ Phys.\ J.\ D} \textbf{\bibinfo{volume}{51}},
  \bibinfo{pages}{173} (\bibinfo{year}{2009}).

\bibitem[{\citenamefont{{M\"u}ller et~al.}(2010)\citenamefont{{M\"u}ller,
  Zhang, Fermani, Chan, Wang, Zhang, Lim, and Dumke}}]{Mueller09}
\bibinfo{author}{\bibfnamefont{T.}~\bibnamefont{{M\"u}ller}},
  \bibinfo{author}{\bibfnamefont{B.}~\bibnamefont{Zhang}},
  \bibinfo{author}{\bibfnamefont{R.}~\bibnamefont{Fermani}},
  \bibinfo{author}{\bibfnamefont{K.~S.} \bibnamefont{Chan}},
  \bibinfo{author}{\bibfnamefont{Z.~W.} \bibnamefont{Wang}},
  \bibinfo{author}{\bibfnamefont{C.~B.} \bibnamefont{Zhang}},
  \bibinfo{author}{\bibfnamefont{M.~J.} \bibnamefont{Lim}}, \bibnamefont{and}
  \bibinfo{author}{\bibfnamefont{R.}~\bibnamefont{Dumke}},
  \bibinfo{journal}{New.\ J.\ Phys.}  (\bibinfo{year}{2010}),
  \eprint{arXiv:0910.2332}.

\bibitem[{\citenamefont{Shimizu et~al.}(2009)\citenamefont{Shimizu, Hufnagel,
  and Mukai}}]{Shimizu}
\bibinfo{author}{\bibfnamefont{F.}~\bibnamefont{Shimizu}},
  \bibinfo{author}{\bibfnamefont{C.}~\bibnamefont{Hufnagel}}, \bibnamefont{and}
  \bibinfo{author}{\bibfnamefont{T.}~\bibnamefont{Mukai}},
  \bibinfo{journal}{Phys.\ Rev.\ Lett.} \textbf{\bibinfo{volume}{103}},
  \bibinfo{pages}{253002} (\bibinfo{year}{2009}).

\bibitem[{\citenamefont{Bean}(1964)}]{Bean64}
\bibinfo{author}{\bibfnamefont{C.}~\bibnamefont{Bean}}, \bibinfo{journal}{Rev.
  Mod. Phys.} \textbf{\bibinfo{volume}{36}}, \bibinfo{pages}{31}
  (\bibinfo{year}{1964}).

\bibitem[{\citenamefont{Brandt}(1996)}]{Brandt}
\bibinfo{author}{\bibfnamefont{E.~H.} \bibnamefont{Brandt}},
  \bibinfo{journal}{Phys.\ Rev.\ B} \textbf{\bibinfo{volume}{54}},
  \bibinfo{pages}{4246} (\bibinfo{year}{1996}).

\bibitem[{\citenamefont{{M\"u}ller et~al.}(2010, under
  review)\citenamefont{{M\"u}ller, Zhang, Fermani, Chan, Lim, and
  Dumke}}]{Mueller10}
\bibinfo{author}{\bibfnamefont{T.}~\bibnamefont{{M\"u}ller}},
  \bibinfo{author}{\bibfnamefont{B.}~\bibnamefont{Zhang}},
  \bibinfo{author}{\bibfnamefont{R.}~\bibnamefont{Fermani}},
  \bibinfo{author}{\bibfnamefont{K.~S.} \bibnamefont{Chan}},
  \bibinfo{author}{\bibfnamefont{M.~J.} \bibnamefont{Lim}}, \bibnamefont{and}
  \bibinfo{author}{\bibfnamefont{R.}~\bibnamefont{Dumke}} (\bibinfo{year}{2010,
  under review}).

\bibitem[{\citenamefont{Buckel and Kleiner}(2004)}]{Kleiner}
\bibinfo{author}{\bibfnamefont{W.}~\bibnamefont{Buckel}} \bibnamefont{and}
  \bibinfo{author}{\bibfnamefont{R.}~\bibnamefont{Kleiner}},
  \emph{\bibinfo{title}{Superconductivity}} (\bibinfo{publisher}{Wiley-VCH,
  Weinheim}, \bibinfo{year}{2004}), \bibinfo{edition}{2nd} ed.

\bibitem[{\citenamefont{Mints and Rakhmanov}(1981)}]{Mints}
\bibinfo{author}{\bibfnamefont{R.~G.} \bibnamefont{Mints}} \bibnamefont{and}
  \bibinfo{author}{\bibfnamefont{A.~L.} \bibnamefont{Rakhmanov}},
  \bibinfo{journal}{Rev. Mod. Phys.} \textbf{\bibinfo{volume}{53}},
  \bibinfo{pages}{551} (\bibinfo{year}{1981}).

\bibitem[{\citenamefont{Hufnagel et~al.}(2009)\citenamefont{Hufnagel, Mukai,
  and Shimizu}}]{Hufnagel}
\bibinfo{author}{\bibfnamefont{C.}~\bibnamefont{Hufnagel}},
  \bibinfo{author}{\bibfnamefont{T.}~\bibnamefont{Mukai}}, \bibnamefont{and}
  \bibinfo{author}{\bibfnamefont{F.}~\bibnamefont{Shimizu}},
  \bibinfo{journal}{Phys.\ Rev.\ A} \textbf{\bibinfo{volume}{79}},
  \bibinfo{pages}{053641} (\bibinfo{year}{2009}).

\end{thebibliography}
\end{document}